\documentclass[useAMS, usenatbib]{mnras}

\usepackage[T1]{fontenc}
\usepackage{ae,aecompl}
\usepackage{graphicx}   % Including figure files
\usepackage{amsmath}    % Advanced maths commands
\usepackage{amssymb}

\usepackage{ulem}
\usepackage{makecell}

\usepackage[dvipsnames]{xcolor}

\usepackage{color}

\title[CLEAN imaging systematics of M87 radio jet]%
{CLEAN imaging systematics of M87 radio jet}
\author[Pashchenko et al.]{\parbox{\textwidth}{
I.~N.~Pashchenko$^{1, 2}$\thanks{E-mail: in4pashchenko@gmail.com},
E.~V.~Kravchenko$^{2,1}$,
E.~E.~Nokhrina$^{2}$,
A.~S.~Nikonov$^{1, 3}$}
\vspace{0.4cm}\\
\parbox{\textwidth}{
$^1$Lebedev Physical Institute, Leninsky prosp.~53, Moscow, 119991, Russia\\
$^2$Moscow Institute of Physics and Technology, Dolgoprudny, Institutsky per., 9, Moscow region, 141700, Russia\\
$^3$Max-Planck-Institut f\"ur Radioastronomie, Auf dem H\"ugel 69, 53121, Bonn, Germany
}
}

\begin{document}

\date{Accepted \dots Received \dots; in original form \dots}

\pagerange{\pageref{firstpage}--\pageref{lastpage}} \pubyear{}

\maketitle

\label{firstpage}

\begin{abstract}
The spectral index images of the jet in the nearby radio galaxy M87 have previously been shown with
Very Long Baseline Interferometric arrays at 2-43\,GHz. They exhibit flattening of the spectra at a location of inner (central) spine and toward outer ridges. This could imply optical depth effects, lower energy cutoff or stratification of the emitting particles energy distribution. In this paper we employ simulations of multifrequency VLBI observations of M87 radio jet with various model brightness distributions. CLEAN deconvolution errors produce significant features in the observed images. For intensity images they result in the appearance of the inner ridge line in the intrinsically edge brightened jet models. For spectral index images they flatten the spectra in a series of stripes along the jet. Another bias encountered in our simulations is steepening of the spectra in a low surface brightness jet regions. These types of the imaging artefacts do not depend on the model considered. We propose a methods for the compensation of the systematics using only the observed data.
% The central spine of M87 radio jet is consistent with optically thin constant spectral index value.
% The intrinsic optically thin spectral index of M87 radio jet is consistent with the constant value.
\end{abstract}


\section{Introduction}
\label{sec:intro}

Active galactic nuclei (AGN) are the most powerful phenomena in astrophysics. The source of an AGN activity, the role of magnetic field, the plasma acceleration and emission of their relativistic jets are being studied for several decades. Generally, it is accepted that plasma in jets is accelerated to the relativistic velocities by magnetic and electric fields, transforming initially highly magnetised outflow into a plasma-dominated one \citep{Beskin06, KBVK07, Lyu09}. However, some questions are still under debate, e.g. the mechanism that accelerates the emitting particles and what is the inner jet structure.  

Radio galaxy M87 is the most famous radio loud AGN due to its proximity and large supermassive black hole (SMBH).
% This makes the M87 the primary target of the Event Horizon Telescope campaign who produced the first image of the SMBH shadow \citep{EHT_I}.
The bright and extended relativistic jet in M87 has been imaged in great details, demonstrating limb-brightening and its transverse stratification \citep{KLH07, 2018ApJ...868..146N}.
% Two and inner ridges
\cite{2016ApJ...817..131H} found that the limb-brightened structure is already well developed at distances $z_{\rm obs} <$ 0.2\,mas at 86\,GHz. 
Using the VSOP dual-frequency observations at 1.6 and 5\,GHz, \citet{2016ApJ...833...56A} observed an additional fainter emission component in the jet centre.
%\cite{2016ApJ...833...56A} used VSOP dual-frequency observations at 1.6 and 5\,GHz and observed three ridges.
This inner ridge line is visible up to 5\,mas from the core in 5\,GHz image. Further down the jet, the central region appears to be dim. At 1.6\,GHz, the inner ridge line becomes visible at 8\,mas up to 16\,mas and, possibly, not well resolved closer to the core. Beyond 20\,mas it is difficult to distinguish three ridgelines.
Signatures of this central ridge were also been observed in another high-resolution images using VLBA + VLA at 15\,GHz \citep{2017Galax...5....2H}, VLBI at 86\,GHz \citet{2018A&A...616A.188K}, and VLBA at 43\,GHz \citep{2018ApJ...855..128W}.
% Previous results on the spectral index
The inner ridge line has initially inverted 1.6-5\,GHz spectral index $\alpha > 1$ close to the VLBI core ($z_{\rm proj} < 5$ mas), that becomes steeper ($\alpha < -1$) beyond 10\,mas from the core \citep{2016ApJ...833...56A}. Hereafter, the spectral index is defined via intensity $\propto \nu^{\alpha}$ and $z$ denotes the distance along the jet.

\begin{figure*}
    \centering
    \includegraphics[width=\columnwidth, trim=0.5cm 1cm 0.5cm 1cm, clip]{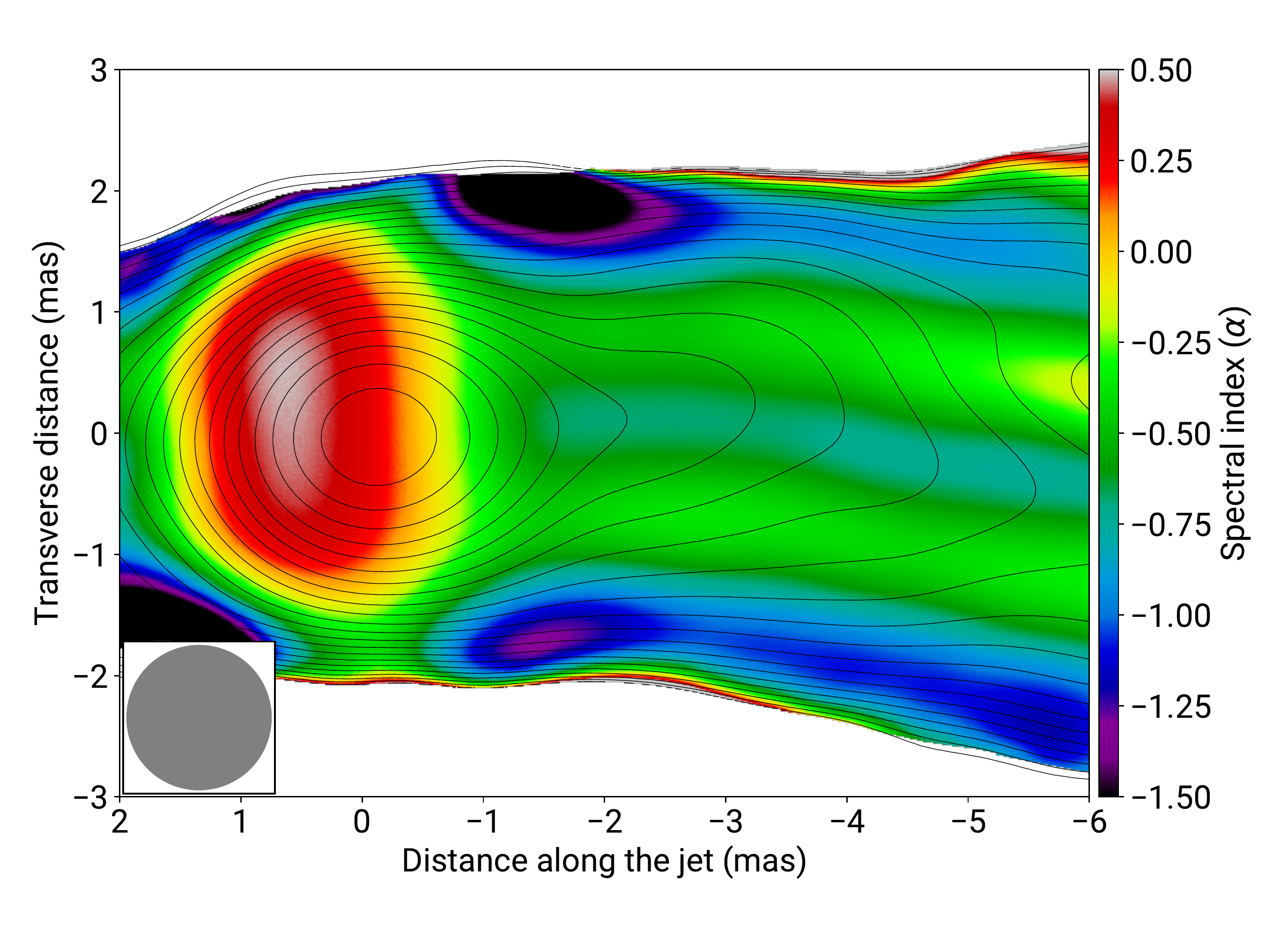}
	\includegraphics[width=\columnwidth, trim=0.5cm 1cm 0.5cm 1cm, clip]{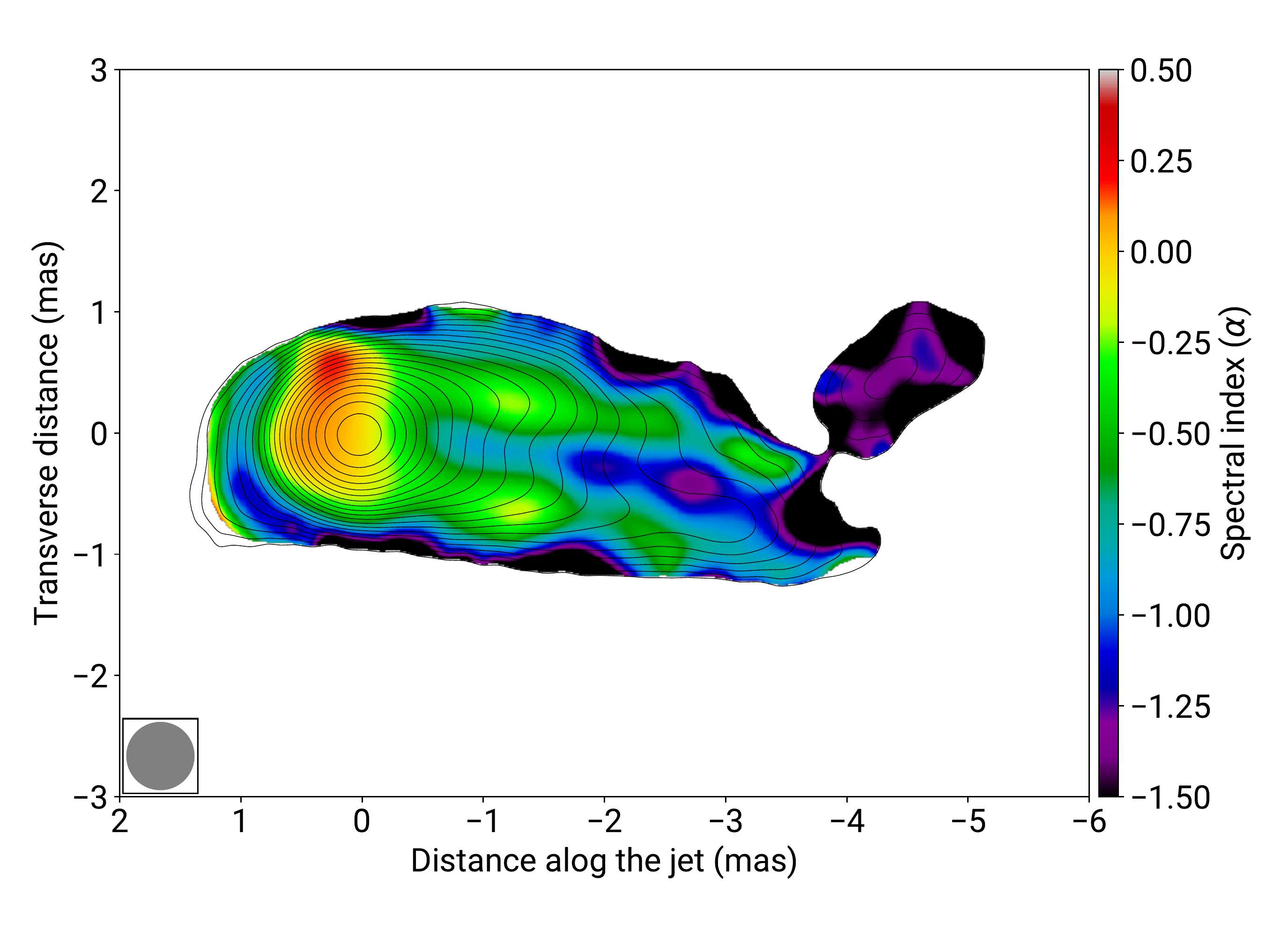}
    \caption{Left: The 8--15\,GHz May 2009 spectral index map (Nikonov et al. in prep.). Contours show the 8\,GHz total intensity image convolved with the circular beam with a size of 1.1\,mas, which equals to equivalent area circular beam of the average of 8 and 15\,GHz elliptical beams. Right: The 24--43\,GHz 18~April~2018 spectral index map \citep{2020A&A...637L...6K}. The contours show the 24\,GHz total intensity image convolved with the circular beam size of 0.5\,mas. }
    % Contours start from 0.5 mJy/beam level at 8~GHz and 0.8 mJy/beam level at 24~GHz increasing by a factor of $\sqrt{2}$.}}
    \label{fig:sp_ind_2243}
\end{figure*}

\citet{2016ApJ...833...56A} discussed two possible origins of the inner ridge line that are consistent with its spectral properties: synchrotron cooling and Doppler deboosting of the fast spine. \cite{2019ApJ...877...19O} employed force-free MHD model and proposed that Doppler boosting due to the MHD velocity field is responsible for the observed central ridge.
Nikonov et al. (in prep.) presented deep 8 and 15\,GHz VLBA $+$ single VLA antenna images of M87 jet, demonstrating three ridgelines. The original most detailed spectral index map reveals spectral flattening both in the centre of the jet and in the pair of outer ridges (see their upper fig.6).
\cite{2020A&A...637L...6K} demonstrated flattening of the spectra in the inner jet using non-simultaneous VLBA observations at 24 and 43\,GHz (see their fig.B.1 and \autoref{fig:sp_ind_2243}, right). The spectra flattens at the two stripes that can be associated with those observed by Nikonov et al. (in prep.) (\autoref{fig:sp_ind_2243}, left). Both images are presented in \autoref{fig:sp_ind_2243} for comparison. At the first sight, this coincidence confirms the significance of this structure.

However, some other multi-frequency VLBI observations of M87 jet do not show such features. \cite{2007ApJ...660..200L} used VLBA and global VLBI data at 22 and 43\,GHz to obtain the spectral index map. It shows constant value $\alpha \approx -0.5$ in the jet with some flattening at the jet edges. The spectral index map of \cite{2019Galax...7...86Z} between 22 and 43\,GHz (their fig. 7a) shows a hint of the outer stripes of the spectral index flattening at first 1\,mas from the core, which was not noticed by the authors. Inverted spectral index at some regions of the jet edges was interpreted as sign of the absorption.

% Why it is important to have an unbiased spectral index maps
Measurements of the spectral properties are essential to better understand the nature of the central and outer ridges of M87 jet.
Flattening of the spectra could imply high opacity to synchrotron self absorption (SSA) or specific emitting particles energy distribution -- with high lower energy cutoff $\gamma_{\rm min}$ or even monoenergetic distribution \citep{1997A&A...328...95B}. However, the lower energy spectra truncation need not to be sharp \citep{2007A&A...463..145T}.
In particular, if the power-law energy distribution of the emitting electrons/positrons has a lower energy cutoff at some $\gamma_{\rm  min}$, then the corresponding spectra flattens below the frequency $\nu_{\rm min}[{\rm MHz}] = 2.8 \gamma_{\rm min}^2 D B[{\rm G}] \sin{\theta}$ to $\alpha_{\rm cutoff}=1/3$ down to $\nu_{\rm SSA}$, where it becomes even more inverted with $\alpha = 2$ \citep[instead of $\alpha = 2.5$ for pure SSA,][]{2012rjag.book.....B}. Here $D$ -- Doppler factor, $\theta$ - LOS angle and $B$ - magnetic field in G.
The lower energy cutoff of the power law distribution is required \citep{2009ApJ...697.1164B} from the modelling the M87 spectral energy distribution (SED)\footnote{\cite{2019A&A...632A...2D} modelled radio (above 43 GHz) to optical jet emission employing relativistic $k$-distribution of electrons. However, their simulations extend up to only $z = 1000 r_{\rm g}$ and, thus, failed to fit radio at a lower frequencies.}. This model predicts the steepening of the originally inverted spectra further down the jet \citep{2007A&A...463..145T,2009ApJ...696.1142M}, as was observed in \cite{2016ApJ...833...56A}. Indeed, as the magnetic field generally declines along the jet, the Lorentz factor $\gamma_{\rm rad}$ of the emitting electrons at a given observed frequency $\nu_{\rm obs}$ increases down the jet as $\gamma_{\rm rad} \approx \sqrt{\nu_{\rm obs}/(D \nu_{\rm B})}$, where $\nu_{\rm B} = eB/(2\pi m_e c)$ -- the Larmor frequency for a given magnetic field. Thus, if in the core region the break frequency $\nu_{\rm min}$ corresponding to $\gamma_{\rm min}$ was higher than $\nu_{\rm obs}$ and the spectrum was inverted, at some distance downstream where $\nu_{\rm min} = \nu_{\rm obs}$ the observed spectrum steepens to its optically thin value $\alpha = -(s - 1)/2$ or even steeper if cooling is important. In the MHD jet models the magnetic field could increase toward the jet axis \citep{Beskin09,Kom09,Lyu09}, that could provide the required high magnetic field value for the spectral flattening due to both SSA and to the lower energy cutoff $\gamma_{\rm min}$.

Particle-in-cell (PIC) simulations of trans-relativistic electron-ion reconnection \citep{2018ApJ...862...80B} showed that electron spectra could be described as a power-law that becomes harder with increasing jet magnetization $\sigma$ (i.e. magnetic field to plasma energy density ratio in a plasma frame) at large plasma $\beta$ (gas to magnetic field pressure ratio).
Recently \citet{2021arXiv211102517C,2021arXiv211102518F} used general-relativistic magnetohydrodynamic simulations and general-relativistic radiative transfer calculations to reproduce the broadband spectrum of M87 from the radio to near-infrared bands and simultaneously fitted the jet collimation profile using the reference frequency of 86\,GHz. They assumed $\kappa$-distribution of electrons and relation between the $\kappa$ parameter, jet magnetization $\sigma$ and plasma $\beta$, estimated by \citet{2018ApJ...862...80B} from their simulations. \cite{2021arXiv211102518F} considered only the emission from the jet ``sheath'' defined by the magnetization cutoff $\sigma_{\rm cut} \propto 1$ and found that the increase of the $\sigma_{\rm cut}$ leads to flatter spectra in the central jet region.
However, they did not simulate the real observables -- interferometric visibilities, but used model images convolved with a Gaussian beam for comparison with the observed images.

Thus, spectral information is essential for studying both the structure of the jet and the particles acceleration process.
Besides the physical processes, systematic effects connected with the sparse ($u$,$v$)-coverage, convolution or imaging algorithm may be responsible for the observed spectral flattening.
\cite{1998A&AS..132..261L} made simulations of VLBA multifrequency observations to assess the accuracy of the spectral index maps, produced with CLEAN algorithm \citep{CLEANref,1980A&A....89..377C}. They found that typical spectral index accuracy is within 90$\%$ for pixels with signal-to-noise ratio SNR > 7 and slightly decreases with distance from the phase centre. However, the systematic part of the uncertainty is to be determined.
\cite{fromm_etal13} showed that when the frequency range is less than a factor of four, the ($u$,$v$)-coverage mainly affects edges of the spectral distribution.
\cite{MOJAVE_XI} made simulations to explore the influence of the uneven ($u$,$v$)-coverage of the multi-frequency data sets on the spectral index maps and found the typical spectral steepening of the spectra in a jet at a level of -0.1.
%Our contribution 
In this paper we use archival dual-frequency VLBI data and simple jet brightness models to address the nature of the spectral flattening in the central region and toward edges of the M87 jet, which we attribute to the imaging artefacts.

The paper is organised as follows.
In \autoref{sec:Model} we describe the model brightness distributions and methods for generating the artificial spectral index maps. In \autoref{sec:Data} we describe the real data sets employed in simulations.
We present simulated spectral index images in \autoref{sec:images} and compare them with the observed maps.
We discuss the systematics in \autoref{sec:bias}, possible ways for its compensation in \autoref{sec:compensation}, list the astrophysical implications in \autoref{sec:implications} and summarise our findings in \autoref{sec:conclusions}. 
We adopt a cosmology with $\Omega_m=0.27$, $\Omega_\Lambda=0.73$ and $H_0=71$~km~s$^{-1}$~Mpc$^{-1}$ \citep{Komatsu09}.

\begin{figure*}
\centering
\includegraphics[width=\linewidth]{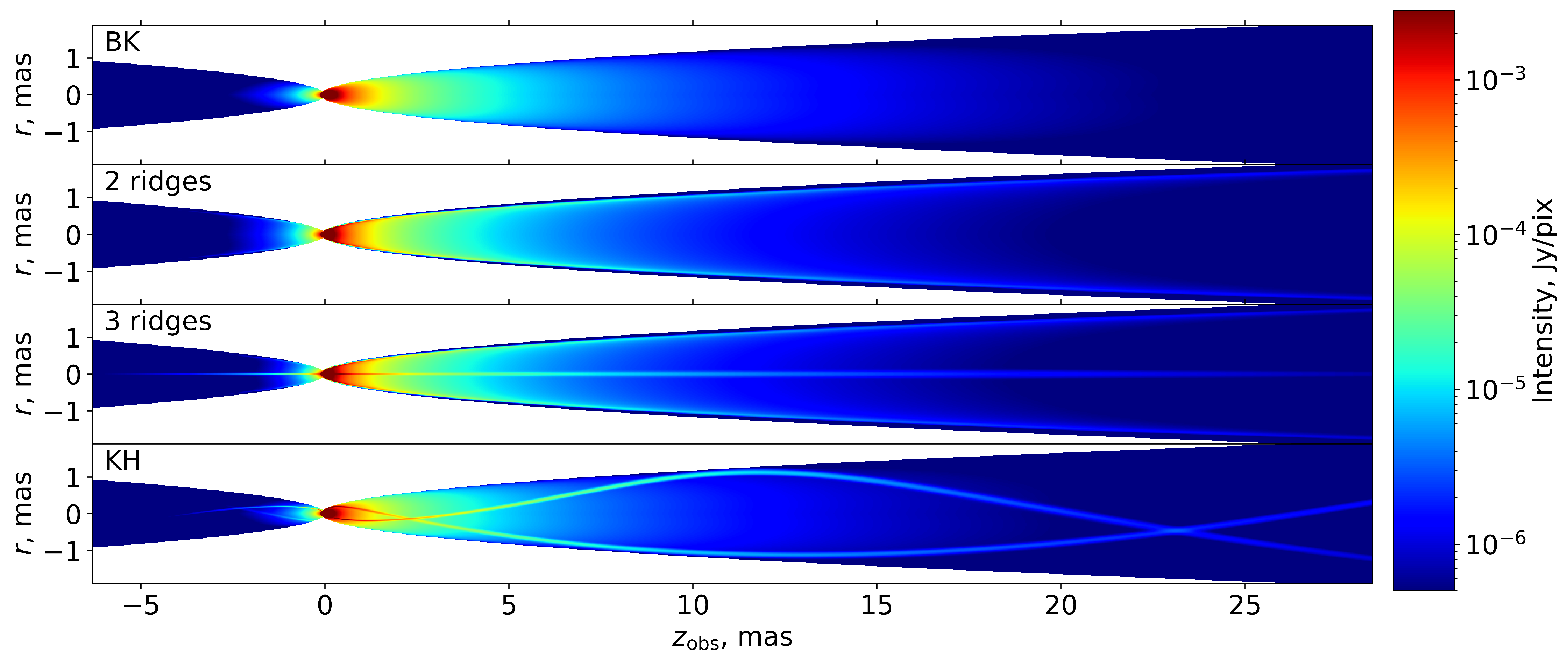}
\includegraphics[width=0.75\linewidth]{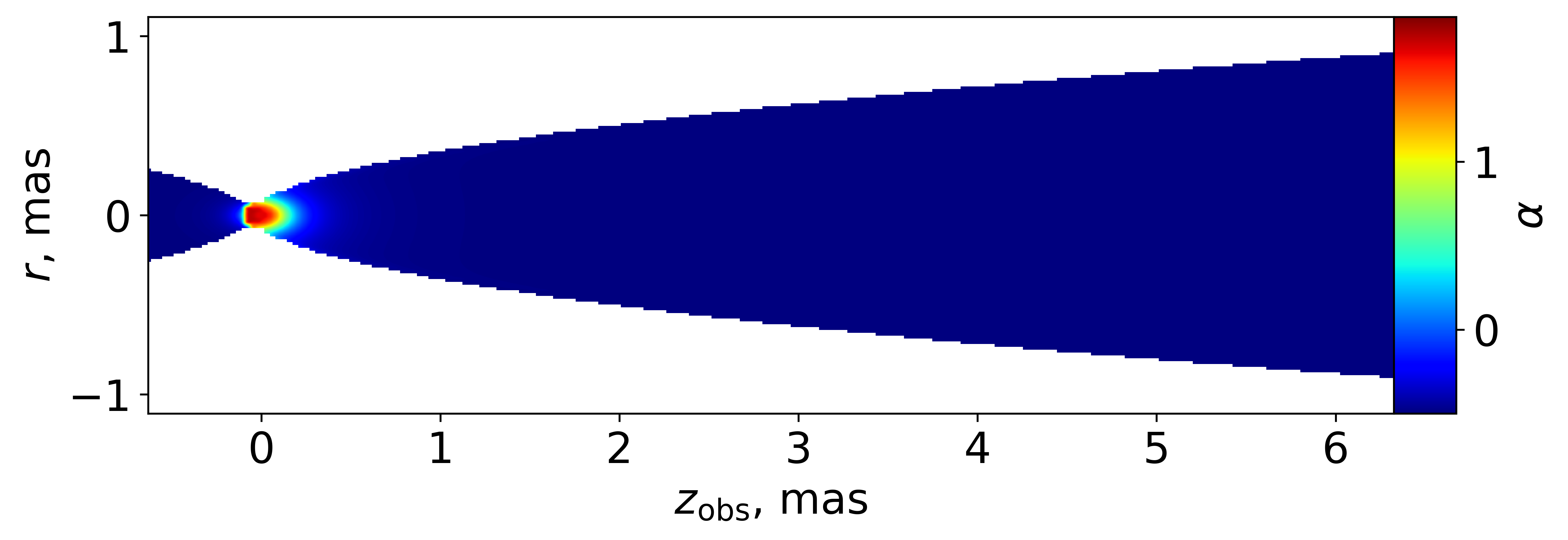}
\caption{\textit{Upper}: The 15\,GHz intensity distributions used for simulations. The models are listed in \autoref{tab:JetModels}. Colour shows the levels in logarithmic scale from $5\cdot10^{-7}$ Jy/pix (dark blue) to 0.1 (red) from the maximum brightness. The full flux is $\approx3$ Jy in all cases. \textit{Lower}: The 8 - 15 GHz spectral index distribution for ``2 ridges'' model. The model optically thin spectral index is $\alpha = -0.5$.}
\label{fig:JetModels}
\end{figure*}

\begin{table}
	\centering
	 \caption{Intensity distribution models used in simulations and plotted in \autoref{fig:JetModels}.}
	\label{tab:models}
	\begin{tabular}{ll}
	 Model type & Heating prescription \\
	 (1) & (2)\\
	\hline
BK & $N(r)=$const uniform across the jet \\
2 ridges &  double humped (edge-brightened) \\
3 ridges &  triple humped (edge-brightened with central spine) \\
KH &  Kelvin-Helmholtz modes \\
 \hline
 \end{tabular}
 \label{tab:JetModels}
\end{table}

\section{The model and simulations}
\label{sec:Model}

To check the possible bias in the VLBI images of the spectral index one has to know the ``true'' model of the spectral index and, thus, the brightness distribution. In general, it depends on the jet magnetic field, plasma bulk velocity field and the distribution of the emitting (i.e. heated) particles. Among these, one of the least certain part of the model is the heating mechanism \citep{2005MNRAS.360..869L,2011ApJ...737...42P,2018Galax...6...31A,BMR-19,2020ApJ...900..100R,2021NewAR..9201610K,2021ApJ...907L..44S}. However, we are interested in the possible instrumental systematics due to imaging and not in the detailed modelling the observed jet structure \citep[e.g. as in][]{2019A&A...629A...4F,2019MNRAS.488..939P}. Thus we decided to build the ``Ground Truth'' brightness model using simple relativistic jet model with ``by-hand'' description of the particle heating and fixed set of the parameters. We consider the M87 jet viewing angle of $\theta=17^{\circ}$ \citep{2018ApJ...855..128W}, parabolic jet geometry $r \propto z^{0.5}$, that is closed to the observed one \citep{Asada12,Hada13}, and selected the radius at $z = 1$ pc ($R_{1{\rm pc}}$) to obtain nearly the same jet width of $\approx 7$ mas (beam convolved) as observed at 15\,GHz at the distance $z_{\rm proj} = 10$ mas (Nikonov et al. in prep.). 
Although the observed jet can be traced up to $z_{\rm proj} \approx$ 500 mas (e.g. in the high-sensitivity 8\,GHz images of Nikonov et al. in prep.), we restricted the modelled jet to the $z_{\rm proj}$= 40\,mas.
This is justified because the observed non-trivial patterns in the spectral index are concentrated within this jet distance and, as we will show later, the systematic effects arise already at these scales. Second, the errors of $\alpha$ increase dramatically at further distances down the jet. We also keep the total flux of $\approx 3$ Jy at 15 GHz in all models considered below. We employed magnetic field $B \propto z^{-b}$ tangled at small scales, where $b = 0.75$ accounts for possible longitudinal component. Emitting particles from the power-law distribution $N(\gamma) \propto \gamma^{-s}$ with $\gamma_{\rm min} = 10$ and exponent $s = 2.0$ (that corresponds to optically thin spectral index $\alpha = -0.5$) were assumed to follow $N(z) \propto z^{-n}$ with $n = 1.5$. 

\begin{figure*}
\centering
\includegraphics[width=0.8\linewidth]{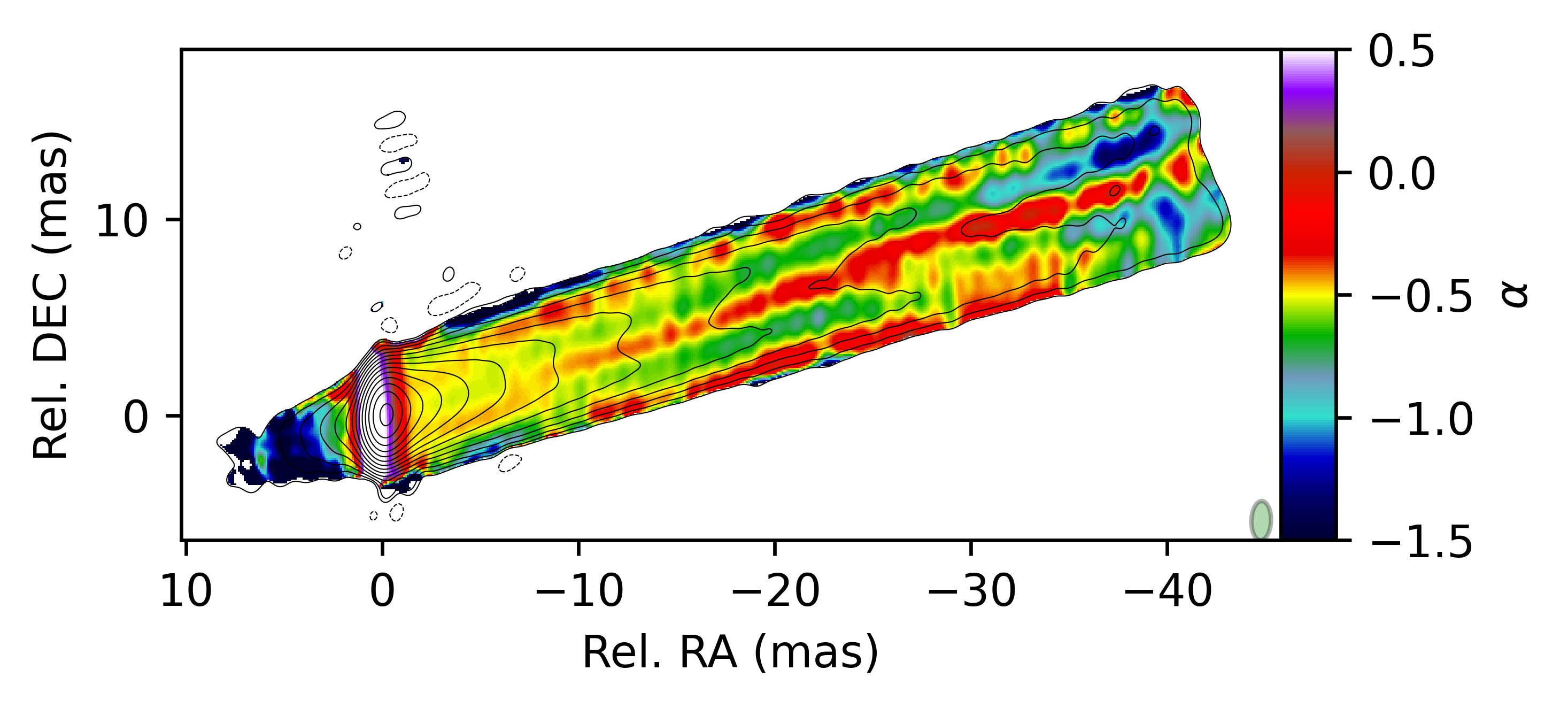}
\includegraphics[width=0.8\linewidth]{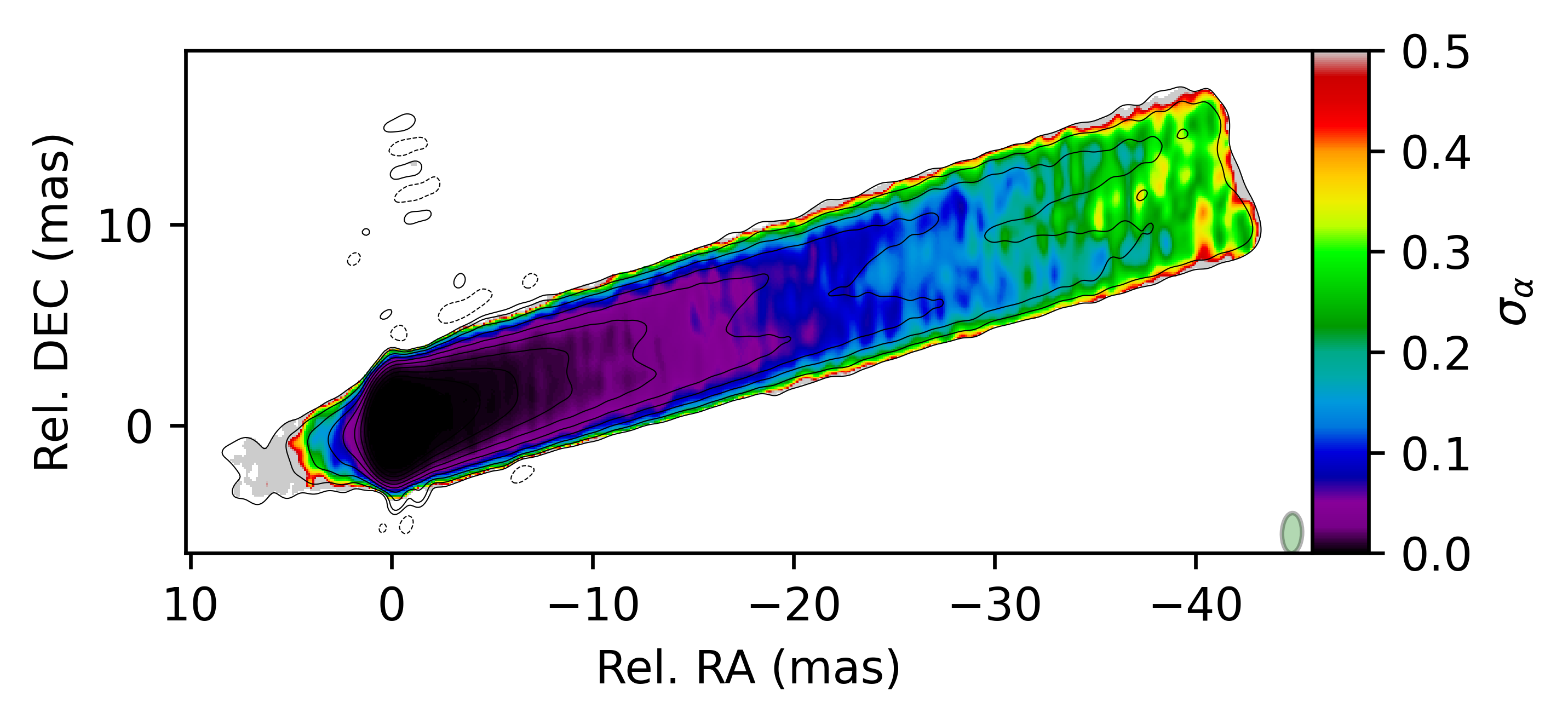}
\caption{The synthetic 8-15\,GHz VLBA + Y1 spectral index map (\textit{Upper}) for ``2 ridges'' jet model with $R_{1{\rm pc}} = 0.1$pc obtained using common 8\,GHz natural weighting beam size (shown at the right bottom corner) superimposed on a Stokes $I$ 8\,GHz intensity contours. The random error of the spectral index (\textit{Lower}).}
\label{fig:alpha_2ridges_bk145_xbeam}
\end{figure*}

We constructed a set of four jet models representing a different transverse brightness distributions: uniform transverse profile (hereinafter -- ``BK'' model, from ``Blandford-K\"{o}nigl''), edge-brightened and edge-brightened with a central spine brightness profiles (``2 ridges'' and ``3 ridges'') and two 3D spirals as a model of the Kelvin-Helmholtz-generated heating (``KH''). The implementation details and physical justifications are presented in \autoref{a:model_physics}.
To obtain the modelled images, we numerically solve radiative transfer equations with synchrotron emission and absorption coefficients as described in \cite{2020MNRAS.499.4515P}. The resulting intensity distributions are summarised in \autoref{tab:JetModels} and their images at 15\,GHz with 8 - 15 GHz spectral index image of ``2 ridges'' brightness distribution are presented in \autoref{fig:JetModels}.
To create a synthetic VLBI image we first Fourier Transformed the modelled brightness distribution to ($u$,$v$) - plane and added the noise, estimated from the scatter of the observed visibilities using the successive differences approach \citep{briggs}. The resulting visibilities were imaged using CLEAN procedure \citep{CLEANref,1980A&A....89..377C} implemented in \texttt{Difmap} package \citep{difmap}. We employed a realization of the multiscale CLEAN algorithm which first uses super-uniform weighting to subtract the compact structure and then a sequence of uniform and, finally, natural weighting for the extended low surface brightness emission. Imaging process stopped when the image \textit{rms}, estimated at the source position equals to the noise \textit{rms}, estimated at 1 angular second away from the source or from the visibility weights. 

Sometimes clipping of the high-frequency data set at the maximum ($u$,$v$)-distance of the low-frequency and at the minimum ($u$,$v$)-distance of the high frequency data set is considered while making spectral index maps \citep[e.g.][]{MOJAVE_XI}. Indeed, the ($u$,$v$)-coverage between different frequencies differs: longer baselines at higher frequencies provide higher resolution, meanwhile lower baselines at lower frequencies better catch the extended emission. This may lead to the artificial spectral index steepening. Thus, we match the ($u$,$v$)-ranges for all data sets, excluding VSOP, where the observed spectral index images were obtained without $uv$-matching.

To obtain the spectral index images for each data set and jet model, we convolved CLEAN components at two frequencies with the same CLEAN beam size corresponding to a low frequency data set. The best beam size in terms of the normalized root-mean-squared error (NRMSE) between the convolved CLEAN model and the true  model is 75-90$\%$ of the CLEAN beam \citep{2016ApJ...829...11C,2017ApJ...838....1A,2018ApJ...858...56K}\footnote{Note, that these NRMSE tests were made with a different model brightness specific to the EHT and NRMSE can not separate systematics from random errors.}. For larger convolving beams the dependence of the NRMSE on the convolving beam is close to the ``ideal'' one. Thus, our choice of the full beam size is safe.

\begin{figure*}
\centering
\includegraphics[width=0.8\linewidth]{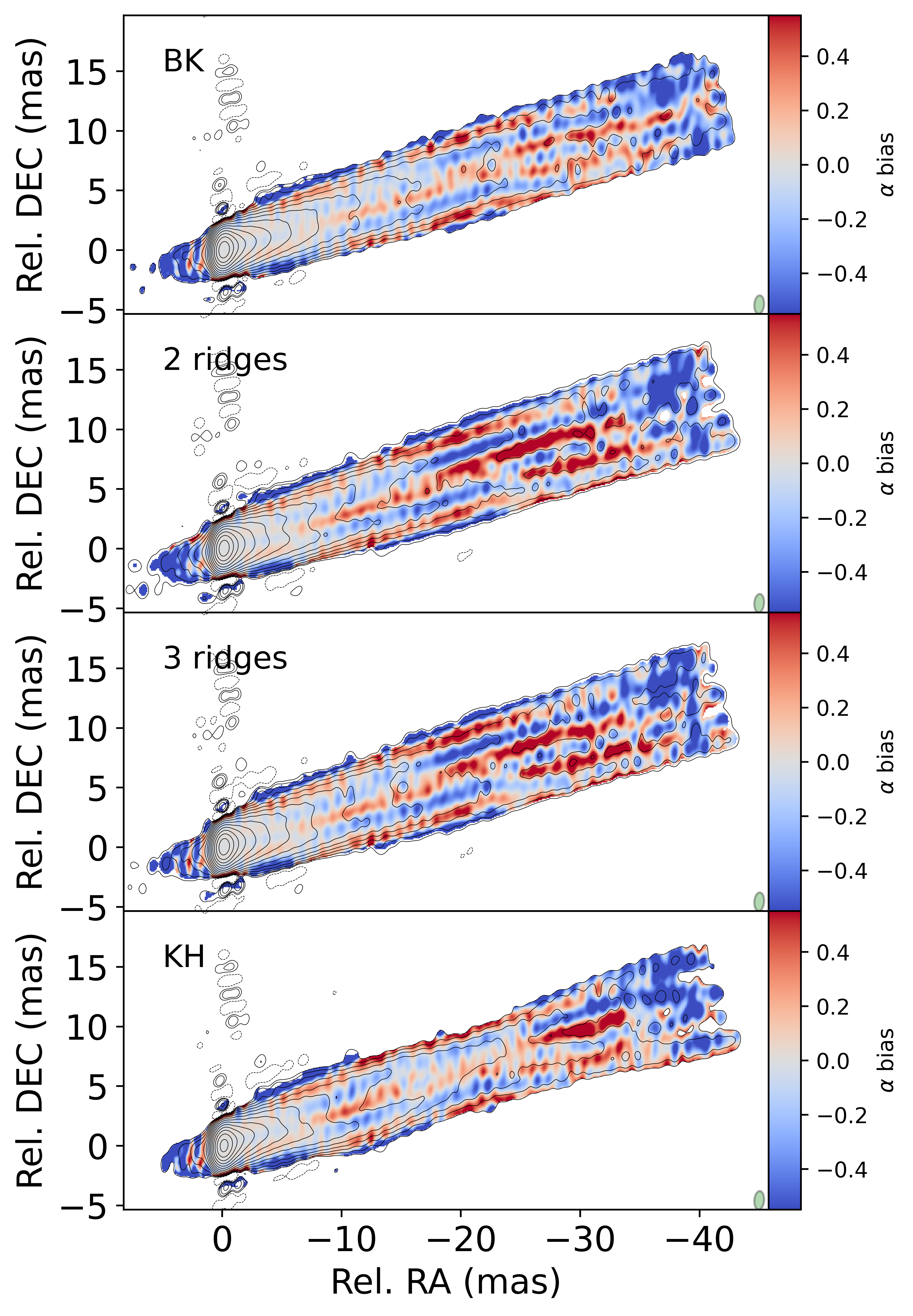}
\caption{The images of the 8-15\,GHz VLBA + Y1 spectral index bias for four jet models with $R_{1{\rm pc}} = 0.12$ pc obtained using common 8\,GHz uniform weighting beam size. Contours represent corresponding 8 GHz Stokes $I$ intensity.}
\label{fig:alpha_bias_bk145_xbeam}
\end{figure*}

\begin{figure*}
\centering
\includegraphics[width=0.65\linewidth]{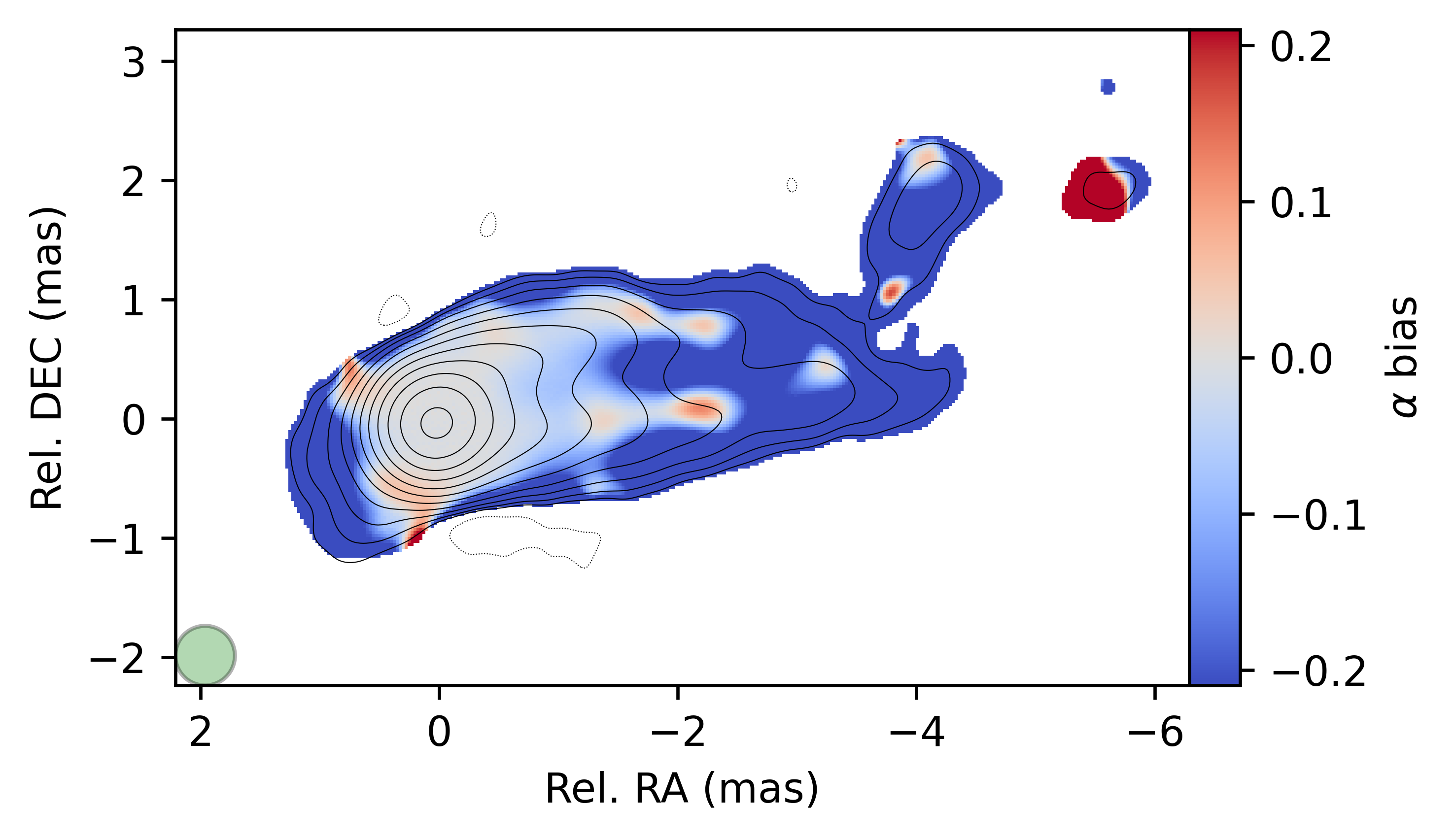}
\caption{Bias estimate of the spectral index for 24 and 43\,GHz VLBA observations based on the original CLEAN models. Circular Gaussian beam size of 0.5\,mas was used.}
\label{fig:SpixBias24_43GHzcirc}
\end{figure*}

\begin{figure*}
\centering
\includegraphics[width=0.7\linewidth]{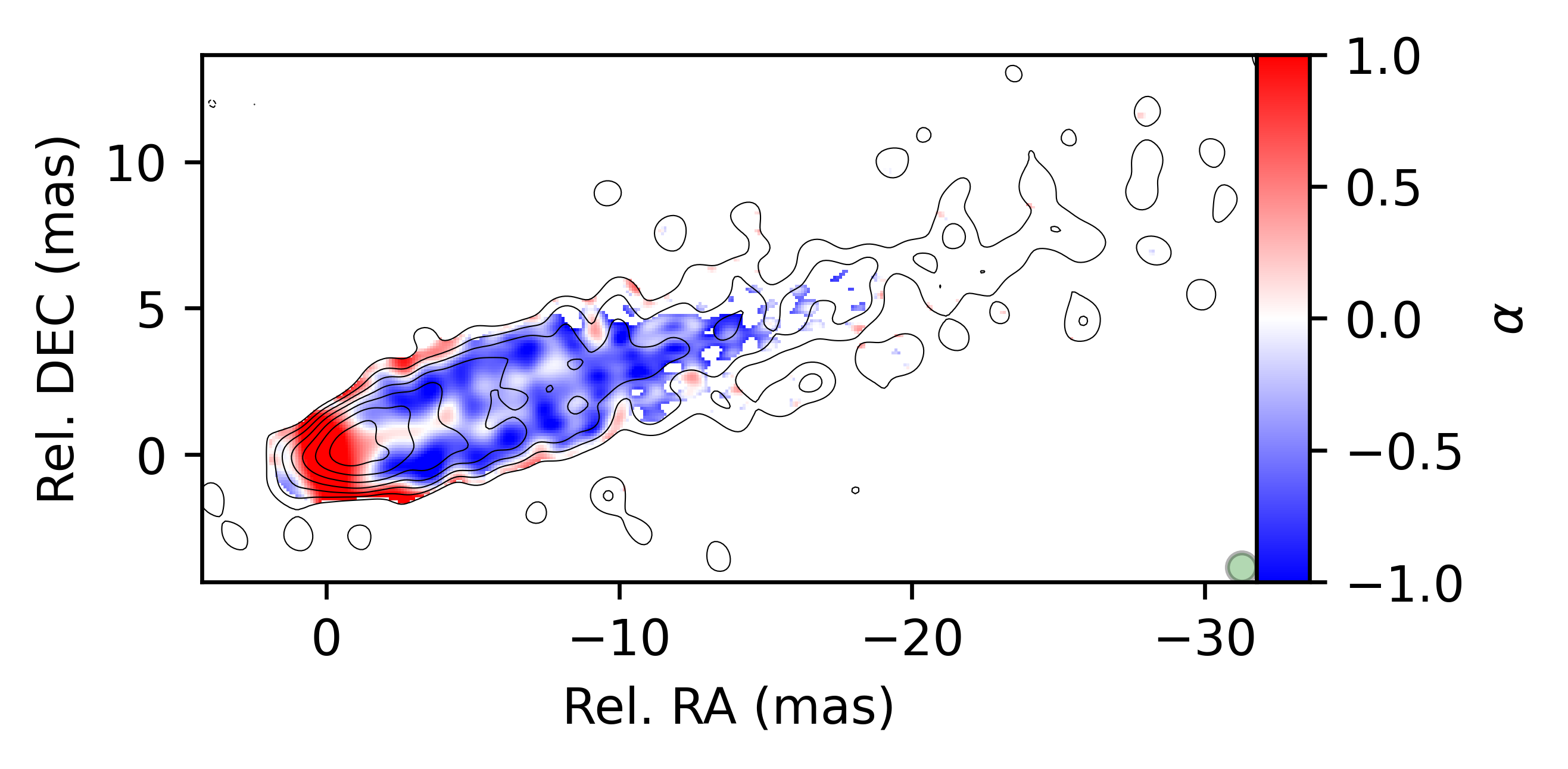}
\caption{Synthetic spectral index map between 1.6 and 4.8\,GHz for the VSOP data set for ``2 ridges'' model brightness with $R_{1{\rm pc}} = 0.12$ pc obtained with 1 mas circular beam. Contours represent the 1.6\,GHz Stokes $I$ intensity.}
\label{fig:ArtSpixVSOP1masbeam}
\end{figure*}

\begin{figure*}
\centering
\includegraphics[width=0.8\linewidth]{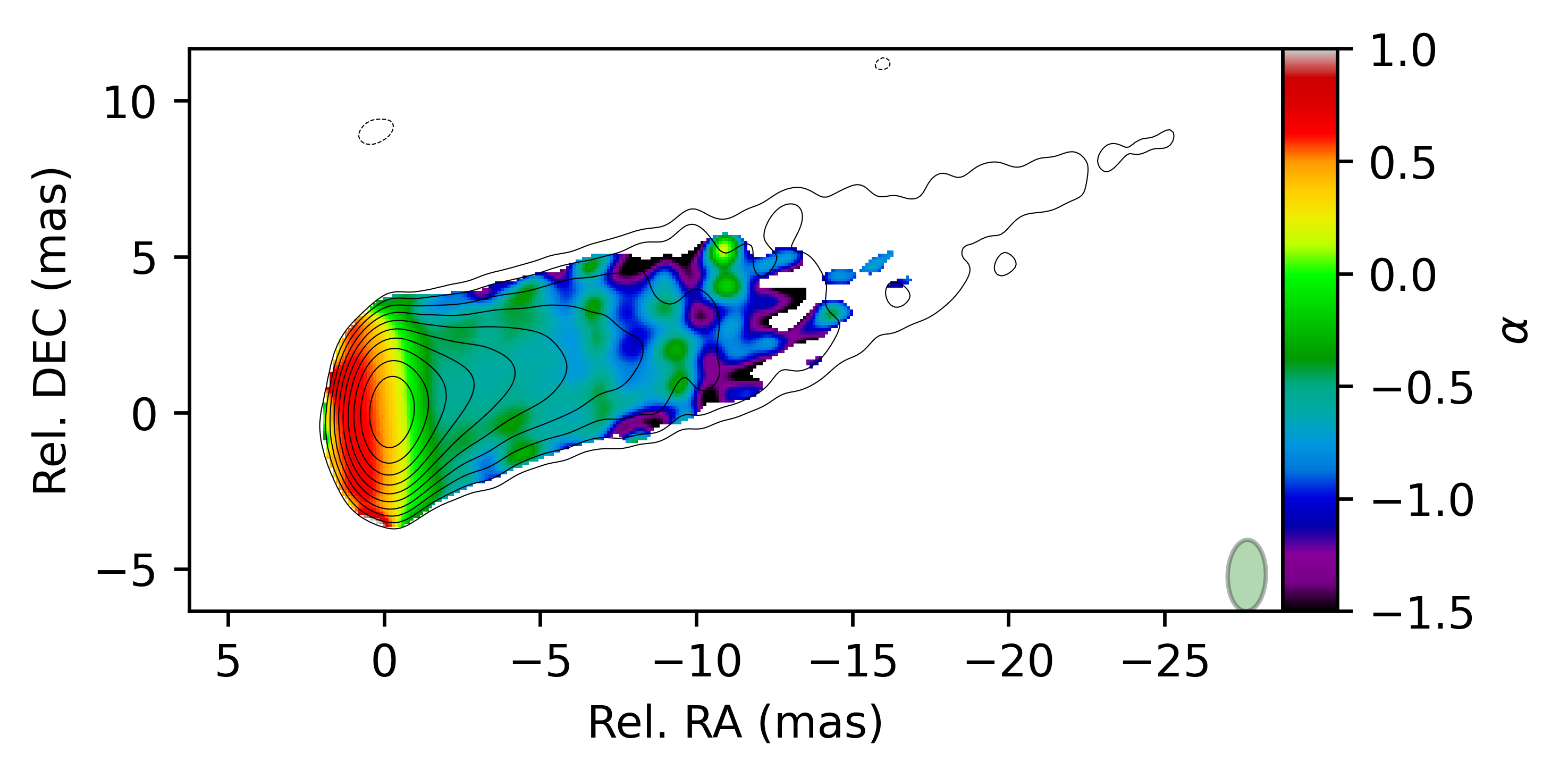}
\caption{Synthetic spectral index map between 8 and 15\,GHz for the MOJAVE data set based on ``KH'' model brightness with $R_{1{\rm pc}} = 0.12$ pc. Contours represent 8 GHz Stokes $I$ intensity and natural weighting beam was used.}
\label{fig:mojave_spectra}
\end{figure*}

\section{Observational data}
\label{sec:Data}

We made use of ($u$,$v$)-coverage of those high-resolution observations, where high-fidelity spectral index images of M87 jet have been presented. This includes 1.6-4.8\,GHz VSOP data \citep{2016ApJ...833...56A}, 8-15\,GHz VLBA data (Nikonov et al. in prep.), 24-43\,GHz VLBA data \citep{2020A&A...637L...6K}. We also considered MOJAVE 8-15\,GHz VLBA data \citep{MOJAVE_XI}.

The imaging VSOP observations of M87 were taken in 2000 March at 1.650, 4.816 and 4.866\,GHz, with the separation of at most three days between the each experiment \citep[project codes W022A7, W040A5, W040A7,][]{2016ApJ...833...56A}. 
Simultaneously with the space radio telescope on board the HALCA satellite, the M87 has been observed by VLBA (1.6 and 4.8\,GHz) together with 70-m telescopes in Robledo and Tidbinbilla (1.6\,GHz).
The signal were recorded in LCP polarization mode with two IFs of 16\,MHz bandwidth.
The achieved resolution was 0.608$\times$0.758\,mas at $-68.2^{\circ}$ at 4.816\,GHz, 0.459$\times$0.853\,mas at $56.6^{\circ}$ at 4.866\,GHz, and of 1\,mas at 1.650\,GHz.
\cite{2016ApJ...833...56A} provided only discussion of the 1.6-4.8\,GHz M87 spectral index.
Earlier study by \citep{2006PASJ...58..243D} that used the VSOP data sets at 1.650 and 4.816\,GHz, demonstrated the pure-quality image only. 

The 8.1-15.4\,GHz spectral index images were presented by MOJAVE based on the quasi-simultaneous observations from 2006 July 15 \citep{MOJAVE_XI}.
The observations were made in dual polarization mode using frequencies centered at 8.104, 8.424, 12.119, and 15.369 GHz. The bandwidths were 16 and 32 MHz for the X and U bands, respectively. The observations were recorded with a bit rate of 128 Mbits s$^{-1}$. In the 8\,GHz bands the observations consist of 2 IFs and 4 IFs in the 12 and 15 GHz bands. 
The spectral index in each pixel was calculated by fitting a power law to the total intensity data using the four frequency bands. 

Global VLBI dual-frequency 8 and 15\,GHz observations of M87 were carried out in 2009 May (project code BK145, Nikonov et al. in prep.). The array included VLBA, single VLA antenna (Y1) and Effelsberg radio telescope, however Nikonov et al. (in prep.) presented spectral index image only for VLBA + VLA(Y1) array. The source was observed during three days in dual polarizaton mode and 16 IFs, with a 8\,MHz bandwidth each.
The synthesised beam is 1$\times$2\,mas at $-2^{\circ}$ at 8\,GHz and 0.6$\times$1.2 at $-10^{\circ}$ at 15\,GHz.

We also used quasi-simultaneous 24 and 43\,GHz VLBA observations discussed in \cite{2020A&A...637L...6K}. M87 was observed in April 2018 in dual polarization mode in eight IFs of 32\,MHz width each (project code BG250A). The resulted beam sizes are the following: 0.76$\times$0.42\,mas at $-9.8^{\circ}$ at 24\,GHz and 0.39$\times$0.21\,mas at $-1.8^{\circ}$ at 43\,GHz.

\section{Simulated spectral index images}
\label{sec:images}
In this section we present the simulated spectral index images for the data sets described in the previous section and compare them with the observed maps.

\subsection{8 -- 15 GHz VLBA (BK145) data}
% \textbf{Common beam:\cite{nikonov} used 1.6x0.8 at PA$= 6^{\circ}$}
As an example of the simulations, in \autoref{fig:alpha_2ridges_bk145_xbeam} we present 8-15\,GHz synthetic spectral index image for ``2 ridges'' brightness model in application to the BK145 VLBA + Y1 data set.
The image reveals clear flattening of the spectra for all four jet models, similar to the observed picture (fig. 6 Upper of Nikonov et al. in prep.).
%The spectral index patterns are similar to the observed picture \citep{nikonov}.
The most prominent patterns are the central and two edge spectral flattening stripes starting from $z_{\rm obs} \approx9$ mas, that are transversely separated nearly by a beam width.
As the jet widens the central flattening splits into a pair of separate stripes at $z_{\rm obs} \approx 25$~mas. In real data this occurs further down the jet at $z_{\rm obs} \approx 50$~mas \citep{nikonov}. This suggests that a real collimation profile differs somehow from the model one.
Less evident, but, as will be shown later, common phenomenon is two parallel stripes of the spectral flattening starting from the core with upper or lower (for BK and KH models) stripe connecting to the central flattening at $z_{\rm obs} \approx5$ mas. The bias of the spectral index for all four models considered is given in \autoref{fig:alpha_bias_bk145_xbeam}. Here we stacked 30 images with different realizations of the thermal noise that were added to the model visibilities to focus on the systematics. Though the flattening pattern is similar for all jet models, edge-brightened models reveal systematics with a larger amplitude comparing to the spine-brightened ``BK'' model. The bias is significant compared to the dispersion (i.e. random error), estimated by the scatter of the spectral index across the different realizations, up to the $z_{\rm obs} \approx 30$ mas, where $\sigma_{\alpha} \approx 0.3$. Note that this error does not include the residual error $\sigma_{\alpha{\rm,sc}} \approx 0.1$ due to uncertainty of the amplitude scale of 5$\%$ at 8 and 15\,GHz. However, this error influences the spectral index at every pixel on the same factor, therefore, it is not relevant for the analysis of the spectral index gradients.

\subsection{VLBA 24 -- 43 GHz data}
% \textbf{CLEAN model as true, common beam: circular 0.5 mas.}
We also tested different types of the true model brightness distribution used in simulations. \autoref{fig:SpixBias24_43GHzcirc} shows the effect of using the original CLEAN model for the 24 and 43\,GHz VLBA data set \citep{2020A&A...637L...6K} instead of the model jet brightness as a true model. This map of the spectral index bias was obtained as the difference between the images of a  simulated spectral index and original spectral index made from the CLEAN components. The common restoring beam size of 0.5\,mas was used, which is 90\% smaller than the original synthesised beam at 24\,GHz. As in the original data (see \autoref{fig:sp_ind_2243}) two stripes of the spectral flattening are evident and nearly coincide with the limb-brightening structure in the total intensity images.

\subsection{VSOP 1.6 -- 4.8 GHz data}
% \textbf{Common beam: circular 1 mas.}
The simulated spectral index map for 1.6 - 5\,GHz VSOP data is presented in \autoref{fig:ArtSpixVSOP1masbeam}, obtained with the common circular beam of 1\,mas size. For the direct comparison, here we used uniform weighting scaled by errors raised to the power $-0.5$ during CLEAN, as was done by \cite{2016ApJ...833...56A}. The central spectral flattening is clearly visible up to $RA = $5\,mas. This position coincides with the position of steepening of the initially inverted spectra of the spine observed in \citet{2006PASJ...58..243D,2016ApJ...833...56A}.

\subsection{MOJAVE 8 -- 15 GHz data}
The simulated two-frequency MOJAVE spectral index image between 8 and 15 GHz is shown in \autoref{fig:mojave_spectra} for ``KH'' brightness model. Here we also stacked 30 realizations of Stokes $I$ images to reduce the variance. Two stripes of the spectral flattening are visible up to $z_{\rm obs} = 5$ mas within the core. 
Interesting that the spectral map obtained by \cite{MOJAVE_XI}\footnote{\url{https://www.cv.nrao.edu/MOJAVE/spmaps/1228+126.2006_06_15_alpha_paper_gray.png}} also demonstrates the hint of two stripes of the spectral flattening up to the same distance. The magnitude of the systematic spectral steepening along the jet observed in out simulations is consistent with the observed one \citep{MOJAVE_XI}, although in general the real spectral index is steeper than simulated. At least partially such systematic offset can be attributed to the residual amplitude scale error.

The Stokes $I$ simulated CLEAN images for both 8 and 15 GHz are shown in \autoref{a:15ghz}.

% Our simulated spectral images also reveal some discrepancies with the observed maps. First, as can be seen from \autoref{fig:alpha_2ridges_bk145_xbeam}, the flattest spectral index in the core region $\alpha_{\rm core,sim} \approx 0.75$ is slightly more inverted than the observed $\alpha_{\rm core,obs} \approx 0.3-0.4$ \citep{nikonov}. The mean core spectral index between 15 and 8 GHz for MOJAVE sample is $\alpha \approx 0.2$ \cite{MOJAVE_XI}, while the median core spectral index between 8 and 2 GHz for a sample of 370 AGNs is $\alpha = 0.3$.

\section{BIAS}
\label{sec:bias}
In previous section we presented the simulated spectral index images for several multifrequency data sets. All of them display stripes of the spectral flattening along the jet or spectral steepening in the outer jet region. As the jet models employed in simulations have constant optically thin spectral index $\alpha = -0.5$ \autoref{sec:Model}, this implies some systematical effect (i.e. bias) involved. In this section we discuss the origin of the bias.

\subsection{The origin of the bias} 
\label{sec:origin}

Experimenting with BK145 data set we found that the systematics that flattens the spectra depends to some extent on the details of the CLEAN procedure. In particular, the systematics is sensitive to the relative length of the uniform and natural weighting steps of the CLEAN procedure. A shallower uniform weighting (i.e. CLEANing up to higher dynamic range before switching to the natural weighting) increases the amplitude of the systematics and slightly changes its morphology. For example, the central flattening becomes connected to the lower stripe of the spectral flattening at $z_{\rm obs} \approx 5$\,mas instead of the upper one.

Bias of the spectral index could result from biased low or high-frequency Stokes $I$ image, or both. We calculated maps of the Stokes $I$ bias for each frequency as the difference between the CLEAN images and the jet model images convolved with the same low-frequency CLEAN beam for BK145 data set. Low frequency image shows significant bias, i.e. much larger (by a factor of $\propto 10$ in $z_{\rm obs} < 10$ mas region) than the random error $\sigma_{I}$. This error was estimated from the scatter of the pixel values between different realizations of the simulated images obtained with different realization of the thermal noise added to the model visibilities (\autoref{sec:Model}). At the same time, the high frequency CLEAN image is practically unbiased: it is dominated by the extended region of a much less significant, i.e. $\propto \sigma_{I}$, negative bias.
As the spectral index $\alpha \propto \log{(I_{\rm high}/I_{\rm low})}$, it is not unexpectedly that the map of the Stokes $I$ bias at 8 GHz perfectly corresponds to the map of the spectral index bias with the opposite sign. 
%
% \nas{Thus, the map of the Stokes $I$ bias at 8 GHz perfectly corresponds to the map of the spectral index bias with the opposite sign, that is expected in case, where a spectral index $\alpha \propto \log{(I_{15}/I_{8})}$.}}

To show the origin of the bias, we considered the distribution of the CLEAN components in a low frequency CLEAN model of the simulated data. CLEAN reconstructs the simulated jet emission by placing CLEAN components preferably in a three ridges at $z_{\rm obs} \gtrsim 10$ mas and in a four ridges at $z_{\rm obs} \gtrsim 35$ mas for jet model with $R_{1{\rm pc}} = 0.12$ pc. These ridges are separated nearly by a beam width. Comparing the spatial distribution of the CLEAN components with Stokes $I$ bias images, it is evident that the flux in the CLEAN map is overestimated between these ridges and underestimated at the ridges. \cite{Cornwell_1983} discussed CLEAN errors, appeared as a multiplicative fringes running through the image with a frequency lying in the un-sampled region of the ($u$, $v$)-plane. They found that fringing is apparent in the error images (i.e. difference from ground truth) but not in the residual images. \cite{1984iimp.conf..255S} (see also \cite{1984A&A...137..159S}) explains the corrugation effect as the result of the negative sidelobes of the beam which cause peaks in the residuals to form at regular intervals in the map. He showed that corrugations correspond to the visibilities outside the range of measurements. However, \cite{10.1093/mnras/220.4.971} showed that the effect can occur even if the beam does not have negative sidelobes or does not occur when the beam has. \cite{briggs} investigated image reconstruction by CLEAN in details and found that it poorly extrapolates. Even downweighted by the restoring beam, the dominant reconstruction errors almost invariably come at or just outside of the ($u$,$v$)-sampling envelope. CLEAN usually makes the extended emission blotchy and may introduce coherent errors such as stripes \citep{Cornwell_1983,1999ASPC..180..151C}. This behaviour of CLEAN can be seen, e.g. in simulations of \citet{2017ApJ...838....1A} (their figure 2), where CLEAN reconstructs the smooth model structure with a patches of the strong components. In our case the jet elongation lines up this patches into several stripes along the jet. The fractional error of CLEAN reconstruction for 8 GHz data set is presented in \autoref{fig:uverror}. Two points are evident.
First, the highest error is located on the outskirts of ($u$,$v$)-coverage with $r_{uv} \approx 100$ M$\lambda$ -- in agreement with simulations of \cite{briggs}. Second, its position angle is perpendicular to the jet, i.e. along the direction of the largest gradient of the model structure. CLEAN model at 15 GHz demonstrates the similar error pattern. However, the longest ($u,v$)-spacing at South--North direction at 15 GHz is almost twice larger than at 8 GHz. Convolution of the 15 GHz CLEAN model with a common 8 GHz beam is equivalent to the multiplication of the CLEAN model Fourier Transform with a Gaussian with FWHM $\approx$ 100 M$\lambda$ in ($u,v$)-domain. This will significantly, almost on an order of the magnitude, downweight the 15 GHz reconstruction error, relative to 8\,GHz.

As can be seen from \autoref{fig:alpha_bias_bk145_xbeam}, the spectral index is biased for all four models considered. This implies that CLEAN introduces biases regardless of the detailed source structure. However, the spectral bias $b_{\alpha}$ is larger for ``2 ridges'' and ``3 ridges'' models. 
Assuming that the bias of the low frequency CLEAN image $b_{I_{\rm low}}$ is relatively small, it is connected with spectral index bias as $b_{\alpha} \approx - b_{I_{\rm low}}/I_{\rm low}$.
Thus, the models with presence of a fine scale structure \autoref{a:model_physics} are more heavily affected by the CLEAN bias. This is a natural consequence of CLEAN error that comes from the outskirts of the ($u$,$v$)-coverage.

\subsection{Influence of the jet -- beam relative orientation}
To explore the influence of the relative orientation of the jet and beam on the bias we simulated BK145 data set with the same $uv$-coverage and noise, but the jet pointing to the South. In that case the resolution is maximal across the jet, while in a real data it is maximal along a jet. The stripes of the artificial spectral flattening remained. However their separation decreased, i.e. there are two stripes of the flattened spectra instead of the central one in the original map \autoref{fig:alpha_2ridges_bk145_xbeam}. The CLEAN reconstruction error reveals the same pattern: the maximal error 5--10$\%$ at the East -- West direction, that is transverse to the jet. 
This result implies that the jet and beam orientations influence the bias differently. Jet orientation determines the direction of the error pattern: in $uv$-plane the CLEAN reconstruction error is maximal at the position angle across the jet, where the brightness gradient is maximal. The maximal $uv$-distance at this position angle, i.e. the beam size in that direction, determines the characteristic angular scale of the error pattern, i.e. the stripes separation.

\begin{figure*}
\centering
\includegraphics[width=0.4\linewidth]{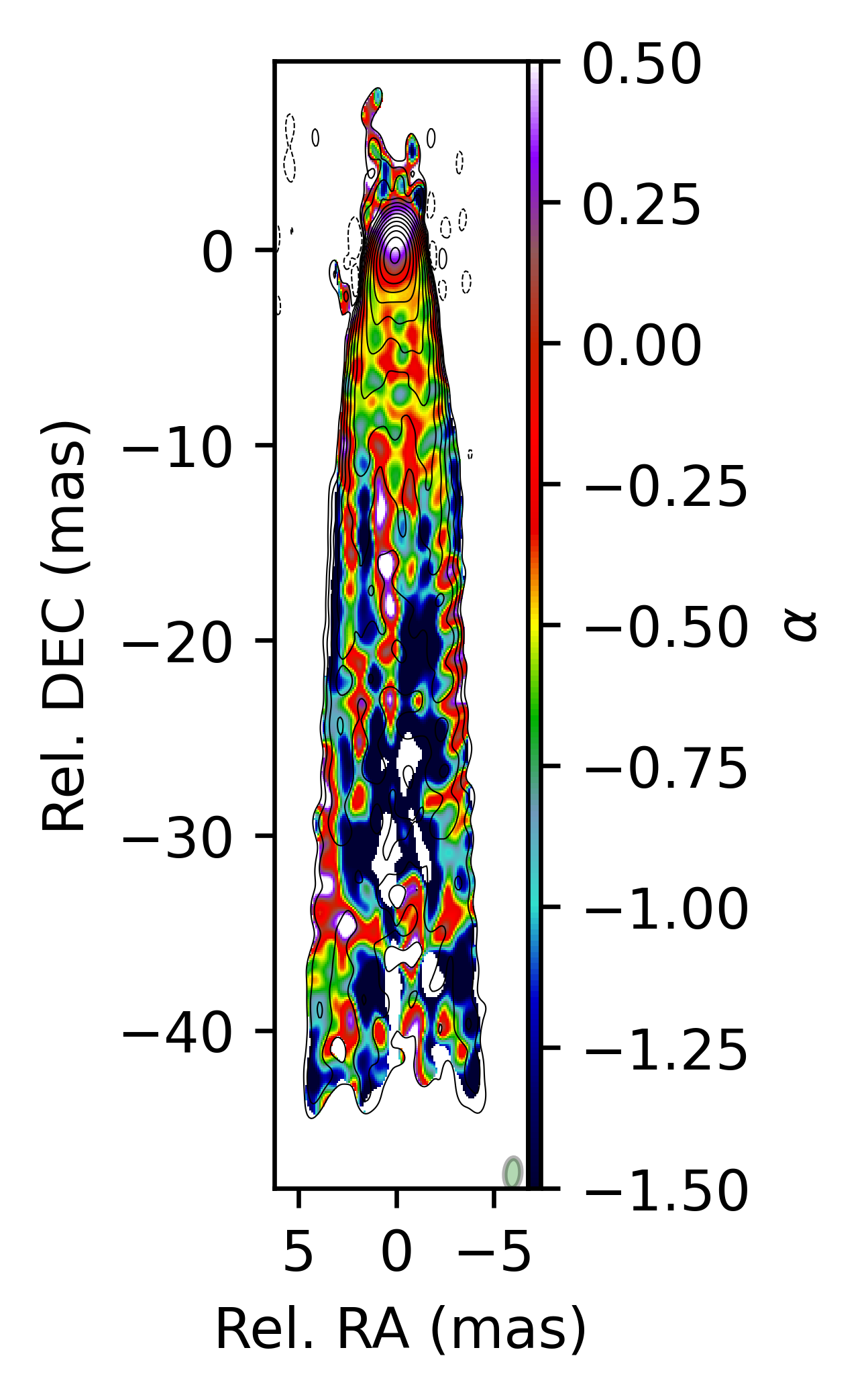}
\includegraphics[width=0.4\linewidth]{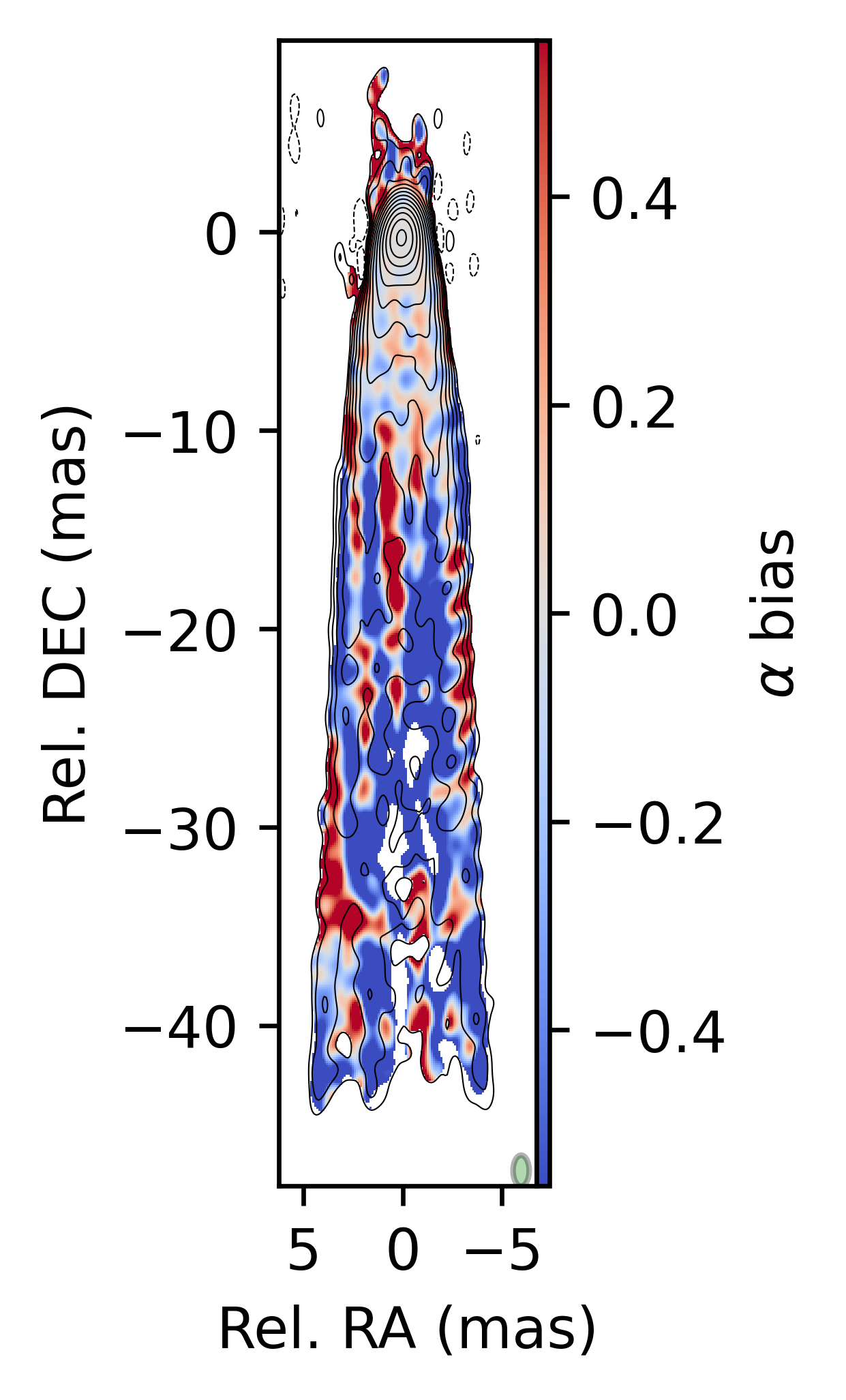}
\caption{Synthetic 8-15 GHz spectral index image (\textit{Left}) and its bias (\textit{Right}) for BK145 data set and ``2 ridges'' model brightness with $R_{1{\rm pc}} = 0.12$ pc obtained with uniform weighting beam, rotated on $\approx90^{\circ}$ relative to M87 jet.}
\label{fig:rot90}
\end{figure*}

\begin{figure*}
\centering
\includegraphics[width=0.48\linewidth]{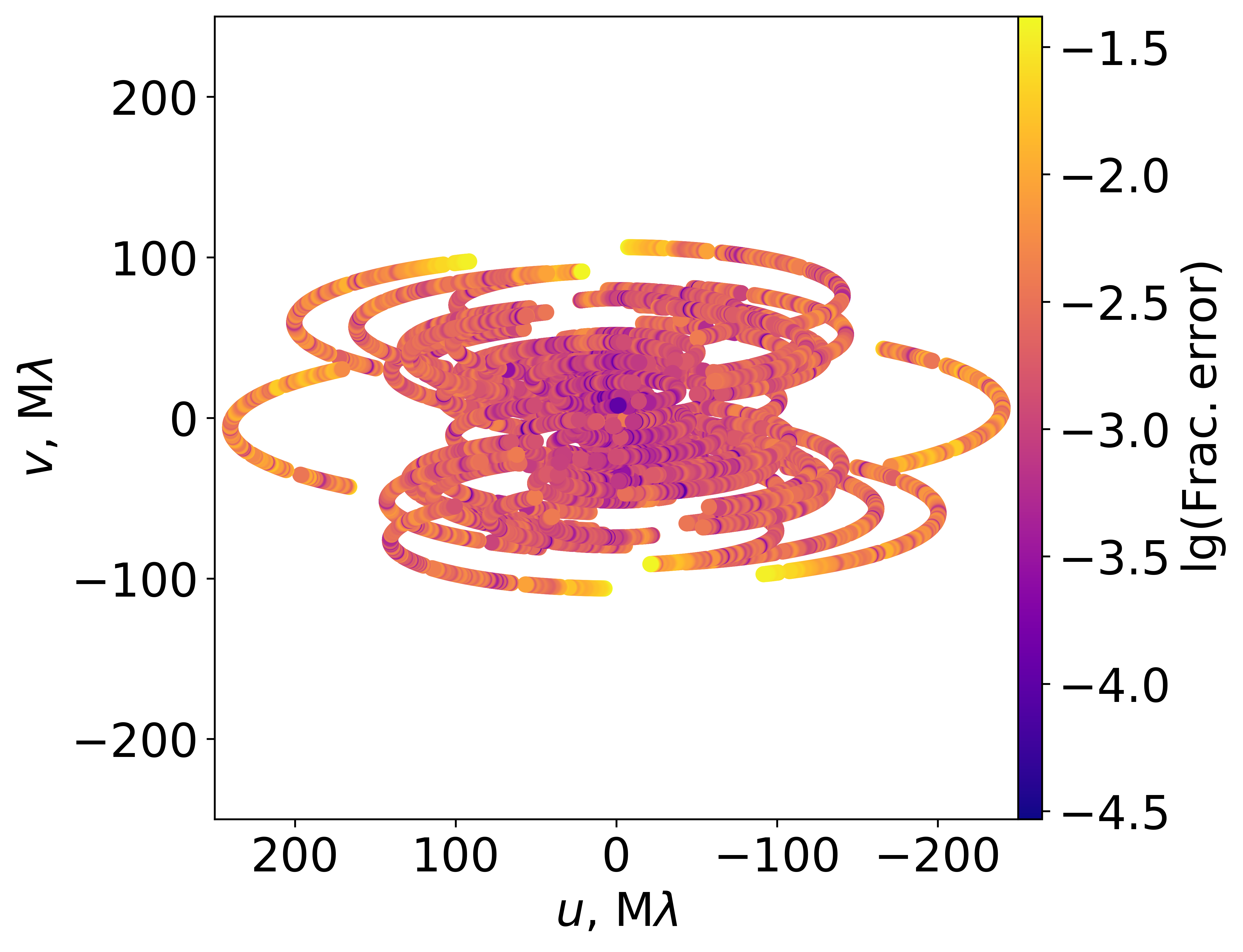}
\includegraphics[width=0.48\linewidth]{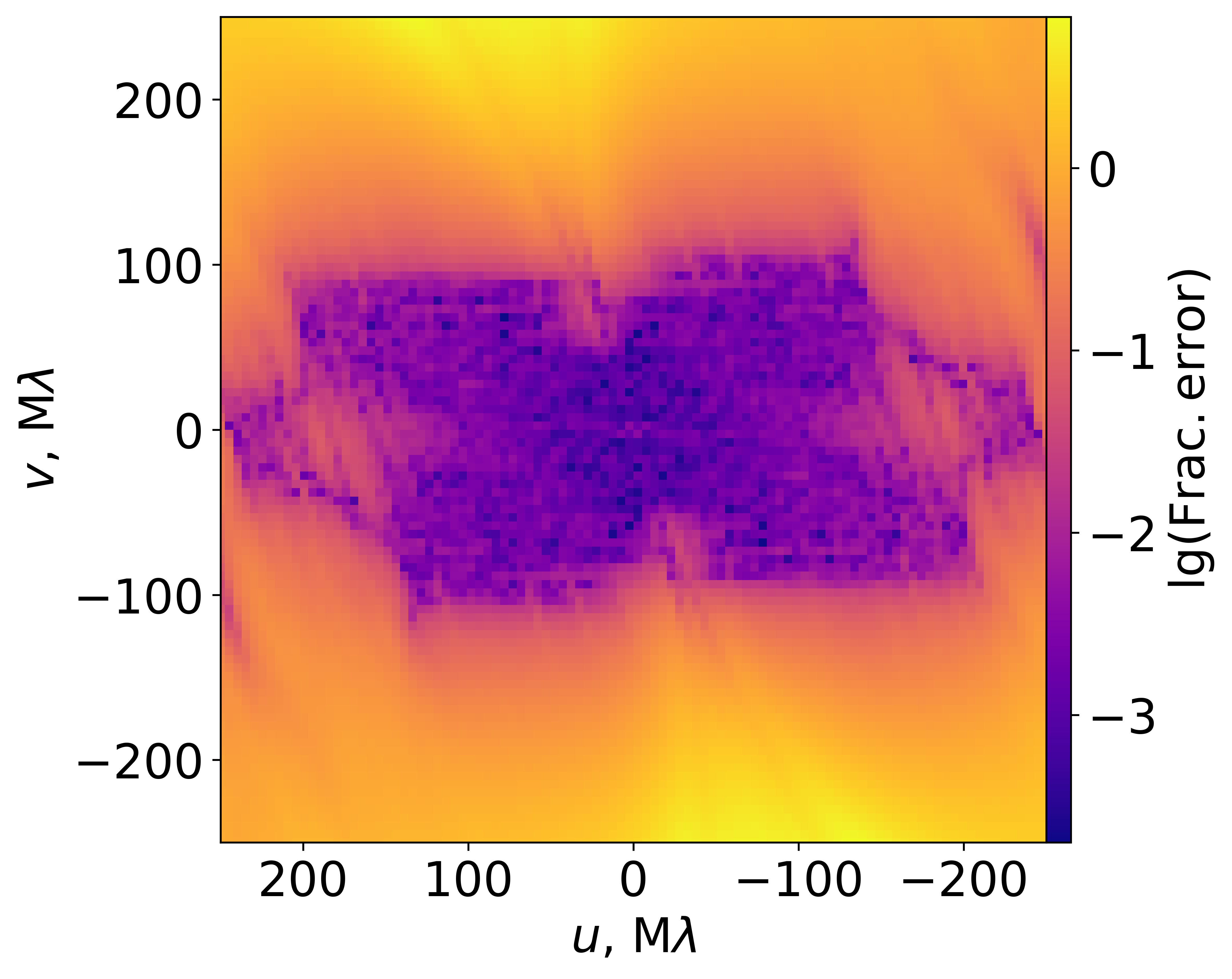}
\caption{Amplitude (logarithm base 10) of the difference between the CLEAN model and the model visibility (``2ridges'') at 8\,GHz for the BK145 data set (\textit{Left}) and in all points of $uv$-plane in a rectangular region (\textit{Right}).}
\label{fig:uverror}
\end{figure*}

\subsection{Influence of the beam size}
\label{sec:beam_size_dep}

To check the dependence of the systematics on the size of the common convolving beam, we repeated our simulations changing the common beam size by a factor ranging from 0.5 to 1.5. Using smaller convolving beams leads to a larger bias and using larger beams decreases the bias. This is the expected behaviour. Indeed, CLEAN reconstruction error itself, i.e. the difference between CLEAN and true model visibilities, does not depend on the size of the convolving beam. It depends on the source structure, $uv$-coverage and weighting scheme employed in imaging. However, larger convolving beams suppress the CLEAN reconstruction error at the largest baselines, where it catastrophically increases \citep{briggs}. Interesting that this mimics the behaviour expected from the real effect, e.g. intrinsic structure in the spectral index distribution. Indeed, larger convolving beams decrease any gradients \citep{MOJAVE_XI,2017MNRAS.468.4478L} and to find the significant small scale gradient one has to increase the resolution -- naively just decreasing the size of the convolving beam.

There is important practical implication of bias dependence on the beam size. It is not an uncommon practice to use the ``equivalent area'' circular convolving beam with size given by $\sqrt{b_{\rm min}b_{\rm max}}$. Here $b_{\rm min}$ and $b_{\rm max}$ -- minor and major axis of the elliptical CLEAN beam. For highly elongated beams this could upweight the CLEAN reconstruction error in the direction, corresponding to the original beam minor axis. The corresponding upweight factor approaches an order of magnitude for elliptical beams with eccentricity $e = 0.2-0.3$ and at the outskirts of the $uv$-sampling envelope.

\subsection{Bias and the inner ridgeline}
\label{sec:bias_inner_ridgeline}

\begin{figure*}
\centering
\includegraphics[width=0.8\linewidth]{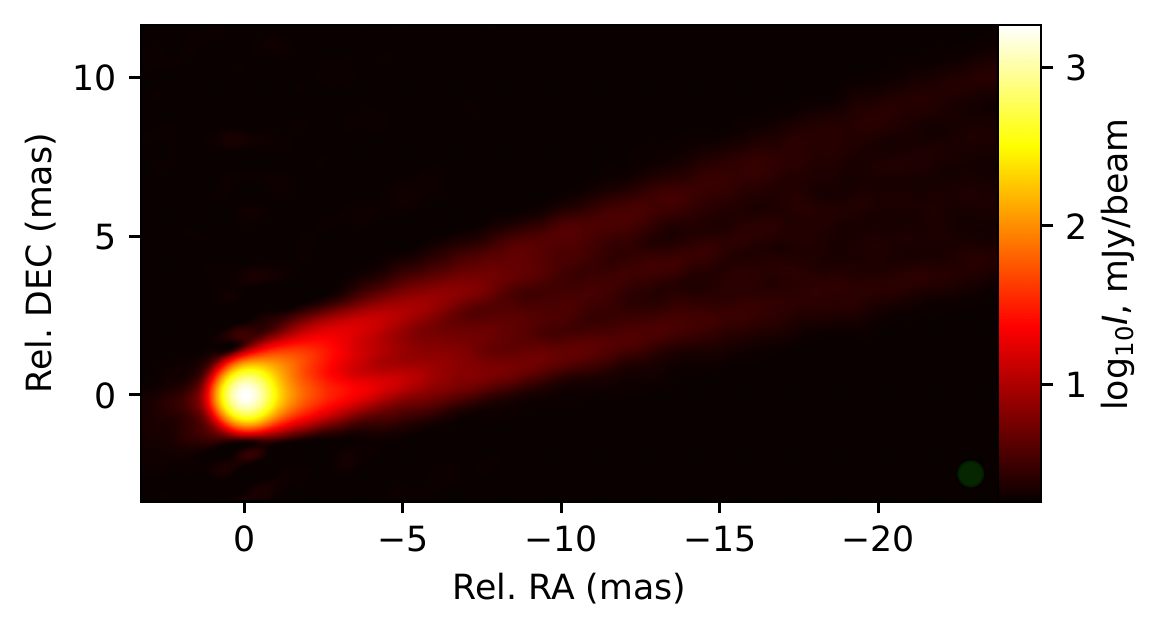}
\includegraphics[width=0.8\linewidth]{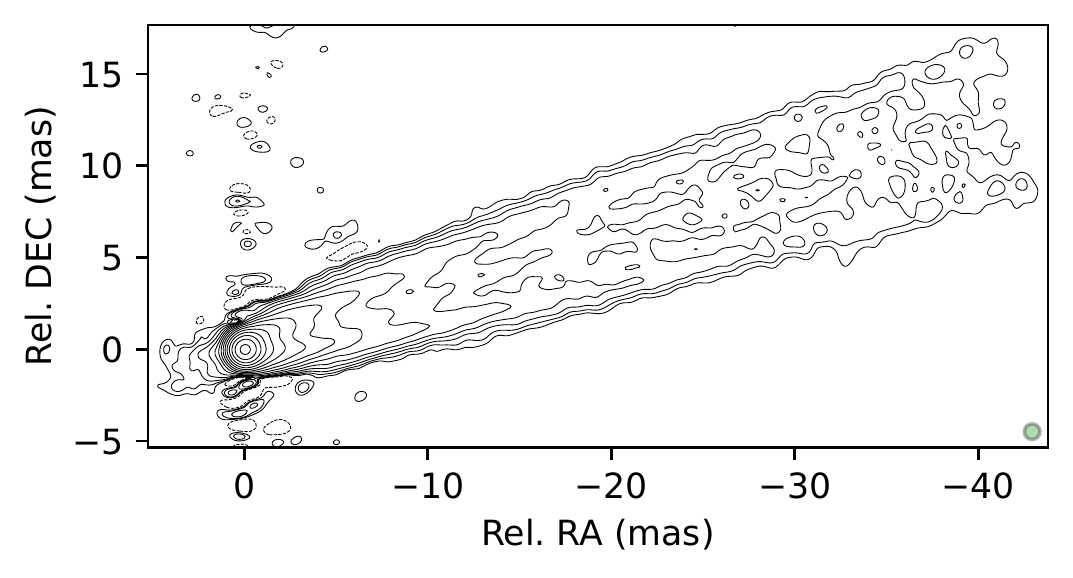}
\caption{CLEAN images of ``2 ridges'' model brightness at 15 GHz made with 0.86 mas circular beam.}
\label{fig:inner_ridgeline}
\end{figure*}

\begin{figure*}
\centering
\includegraphics[width=0.8\linewidth]{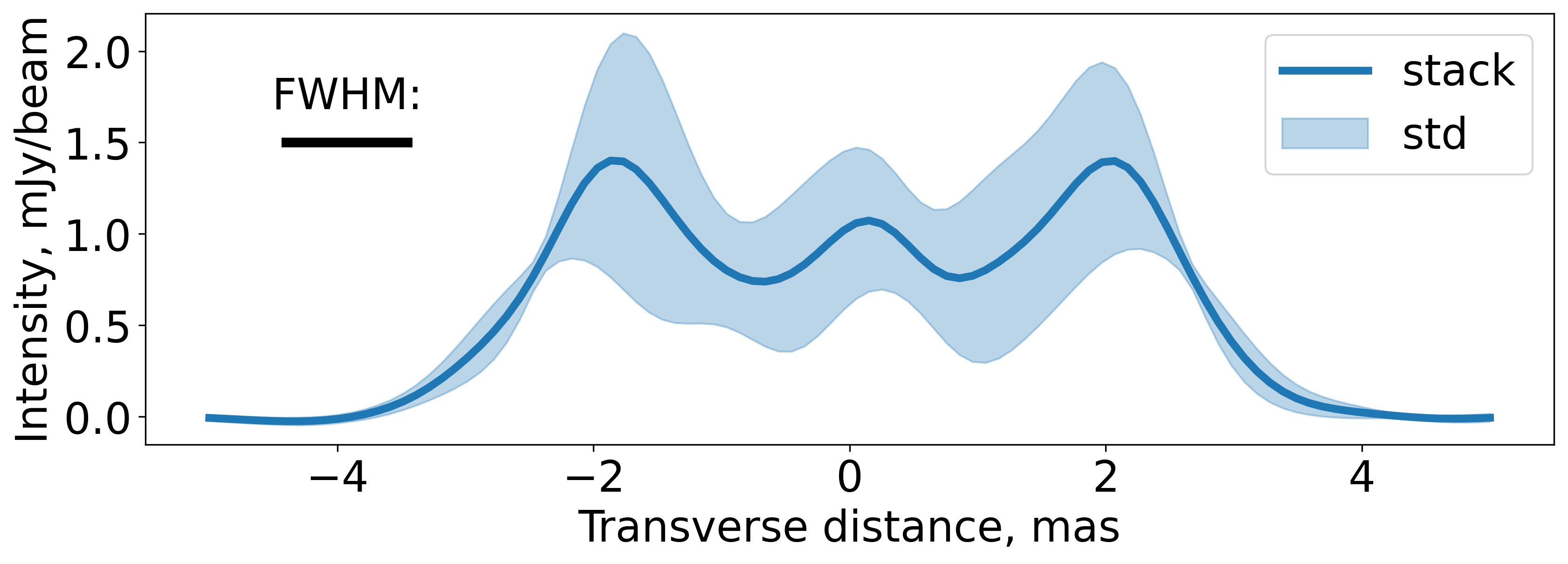}
\caption{Slice across the synthetic 15 GHz CLEAN image presented in \autoref{fig:inner_ridgeline} averaged between 10 and 20 mas. Blue colour band represent standard deviation of individual slices around the mean. Horizontal thick black line represent FWHM of the convolving beam.}
\label{fig:inner_ridgeline_slice}
\end{figure*}

The inner ridge line that was noticed at several frequencies in both Ground only and Space-Ground observations (\autoref{sec:intro}) could be connected with the imaging systematics as well. Nikonov et al. (in prep.) demonstrated prominent central brightening at 15 GHz (their upper fig. 1 and fig. 3). CLEAN image presented in their fig. 4 made use circular 0.86 mas convolving beam, that is superresolved relative to natural weighting 1.2 mas beam size at the direction across the jet. This could increase the bias (\autoref{sec:beam_size_dep}). To check the possible systematics we made CLEAN image of the ``2 ridges'' brightness model \autoref{sec:Model} and used 0.86 mas circular CLEAN beam. The result is presented in \autoref{fig:inner_ridgeline} in both colours and contours. The inner ridge line is clearly visible. Its length depends on the model jet width: in \autoref{fig:inner_ridgeline} with $R_{\rm 1pc} = 0.12$pc it splits in a pair of ridges at RA $= 20$ mas, while in narrower jet with $R_{\rm 1pc} = 0.09$pc it stays prominent up to RA $\approx30$ mas. The slice across the CLEAN image averaged between 10 and 20 mas is presented in \autoref{fig:inner_ridgeline_slice}. is in good agreement with the observed one (fig.3 of Nikonov et al. in prep.). The appearance of the inner ridge line is also apparent, although less clearly, in the 15 GHz CLEAN images convolved with a native 15 GHz CLEAN beam. \autoref{fig:bk145_15GHz_CLEAN_allmodels} presents CLEAN images for all models considered.

\subsection{Spectral steepening along the jet}
\label{sec:steepening}
We also found that the steepening of the spectra is mostly determined by the negative bias in the high frequency Stokes $I$ CLEAN image. This bias depends on the pixel SNR and, thus, grows in the extended jet regions. It also increases with the increasing of the convolving beam size above the original value determined by the dirty beam specific to a given frequency. A spectral index map is usually constructed using the high-frequency image convolved with a lower-frequency beam. Thus, one expects larger value of negative bias for the high-frequency Stokes $I$ image. Note that this is the opposite to the spectral flattening systematics, where the increase of the convolving beam decreases the bias \autoref{sec:steepening}. This implies the different nature of the steepening bias. As we demonstrate further in \autoref{sec:compensation}, CLEANing deeper decreases this systematics. It suggests that the origin of the bias is due to the image residuals. Indeed, any CLEAN image is made of the CLEAN components, convolved with a CLEAN beam and the residuals that are effectively convolved with a dirty beam native to a given frequency \citep{briggs,1999ASPC..180..301F,Rich_2008}. For resolution matching, the high frequency CLEAN components are convolved with a larger (typically low frequency) beam. This effectively decreases the relative magnitude of the residuals and leads to a flux underestimation for a high frequency CLEAN image  This is especially prominent in a low-SNR extended jet regions, where the spectral steepening is typically observed. 

Interesting, that \cite{2023ApJ...945...40C} recently compared the multi-frequency regularized maximum likelihood imaging approach and CLEAN in obtaining the spectral index image of S5 0212$+$73 and NRAO 530. They used 8.1, 8.4 and 12.1 GHz data from \cite{MOJAVE_XI} and found that the recovered spectral index far from the core is generally more negative in CLEAN reconstruction.

\subsection{Connection with calibration errors}

\cite{briggs} found that deconvolution process itself can contribute an error comparable to that of residual calibration errors even for the simplest imaging problems. Moreover, the morphology of deconvolution and calibration errors are similar. For example, short time baseline-dependent (i.e. non-closing) amplitude error causes corrugation in the image plane, with direction corresponding to the position angle of the error in $uv$-plane and characteristic scale inversely proportional to its $uv$-radius \citep{1999ASPC..180..321E}. However, it is unlikely that the residual calibration error has both short time scale and located at the specific position in $uv$-plane, corresponding to jet position angle. Thus, imaging artefacts elongated along the jet are are most likely due to deconvolution errors.

The principal difference between the calibration errors and CLEAN errors is that the former are localised at the sampled $uv$-points, while CLEAN errors keep rising outside of the $uv$-sampling envelope. This makes possible to mitigate the effect of, e.g. gain amplitude error on image fidelity, by flagging the data points affected by the calibration error. 

In our simulations \autoref{sec:Model} focused on the deconvolution performance no residual calibration errors, e.g. gain errors or non-closing errors, are assumed. Thus the imaging performance on real data sets may be worse than given in our simulations. Moreover, the errors introduced by the interaction of the residual calibration errors and deconvolution errors can be worse than the sum of the individual errors \citep{briggs}. This situation is much more complex because calibration errors could have different nature and different appearance in image plane \citep{1999ASPC..180..321E}. However, the situation discussed in our work must be considered as a fundamental ``baseline'' level for a fidelity of the CLEAN images of M87 radio jet.

\section{Bias compensation}
\label{sec:compensation}

In a previous sections we found that spectral index images are affected by the CLEAN systematics. In this section we answer the following question: could the found systematics be compensated using only the observed data? In that case there is no ``true'' model at hand to estimate the bias directly. As the spectral index $\alpha \propto \log(I_{\rm high}/ I_{\rm low})$ and the bias of $\alpha$ is due to bias of $I_{\rm low}$ (\autoref{sec:origin}), there are at least three possible ways to accomplish this task. The first two are to estimate the bias of $I_{\rm low}$ or $\alpha$ using some surrogate of the ``true'' model and correct the observed data with that estimate. The general idea behind this approach is that the observed data (i.e. observed visibilities and corresponding CLEAN maps) is to the true unknown model as the simulated data is to some surrogate of the true model.

\begin{figure*}
\centering
\includegraphics[width=0.75\linewidth]{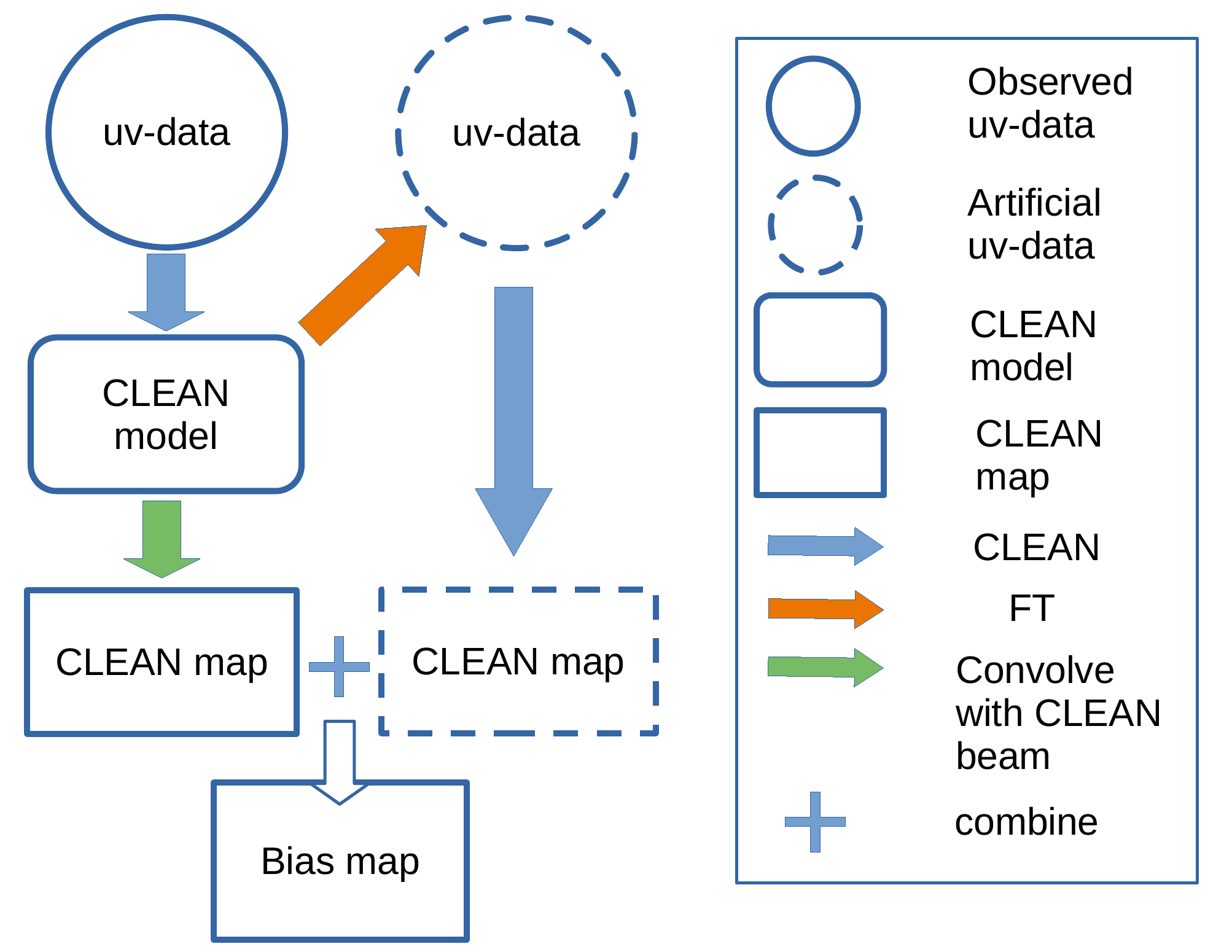}
\caption{Block scheme representing the procedure of the bias estimation using CLEAN model as a surrogate of the true model. \autoref{sec:correcting}.}
\label{fig:scheme_bias_estimation}
\end{figure*}

\subsection{Estimating and correcting the bias in low-frequency or spectral image}
\label{sec:correcting}
For example, CLEAN model of the observed data could be employed as a surrogate of the true model. The general block scheme representing this procedure of bias estimation is presented in \autoref{fig:scheme_bias_estimation}. In our case the jet brightness model \autoref{sec:Model} was considered as a ``truth'' that is unobserved in practice. The \textit{first level} artificial $uv$-data sets, generated from this model at 8 and 15 GHz were considered as ``observations'' and their CLEAN models were used to generate many realizations of the BK145 data set. These \textit{second level} artificial data sets were imaged in the same way, as the original first level artificial data set. Difference between the mean of the obtained Stokes $I$ images and the first level CLEAN model convolved with the same beam was considered as a bias estimate for a given frequency. After correcting both 8 and 15\,GHz first level CLEAN images for the obtained bias estimates the spectral index image was created. To measure how the procedure succeeded with bias compensation we constructed the bias of the obtained spectral index map now using the spectral index image built from the model brightness \autoref{sec:Model} that generated the initial first level artificial ($u$,$v$)-data set. The resulting bias image still displays stripes of the spectral flattening, with a slightly suppressed value and the apparent stripes of the flatten spectra being more patchy. It also shows no negative bias in the counter jet.

We slightly altered the above procedure directly correcting for the bias the spectral index image itself, rather than Stokes $I$ images at each frequency. In this case, the difference between the mean spectral index map across the second level realizations and the spectral index map, obtained from the first level CLEAN models was employed as a bias estimate. Correcting the CLEAN bias in this way affects the spectral index map almost the same as the Stokes $I$ bias compensation. However, this also removes the negative bias from the outer jet region.

Thus, bias correction using its estimate from the observed data, i.e. CLEAN models, is unable to remove the artificial stripes of the spectral  flattening. This could be due to the relatively short length scale of the error pattern.
% Indeed, as CLEAN extrapolates poorly, the result of the extrapolation should heavily depend on the source structure being imaged.
Indeed, as the maximal CLEAN reconstruction error comes at ($u$,$v$)-sampling envelope, the CLEAN prediction at this region should heavily depend on the source fine structure being imaged. In other words, it differs for the visibilities obtained from the real brightness distribution and from the set of CLEAN-components fitted to the real brightness distribution. Moreover, CLEAN-model employed in the bias estimation procedure is itself the result of the CLEAN extrapolation.

\subsection{Re-creating error pattern at high-frequency image}
\label{sec:recreating}

\begin{figure*}
\centering
\includegraphics[width=\linewidth]{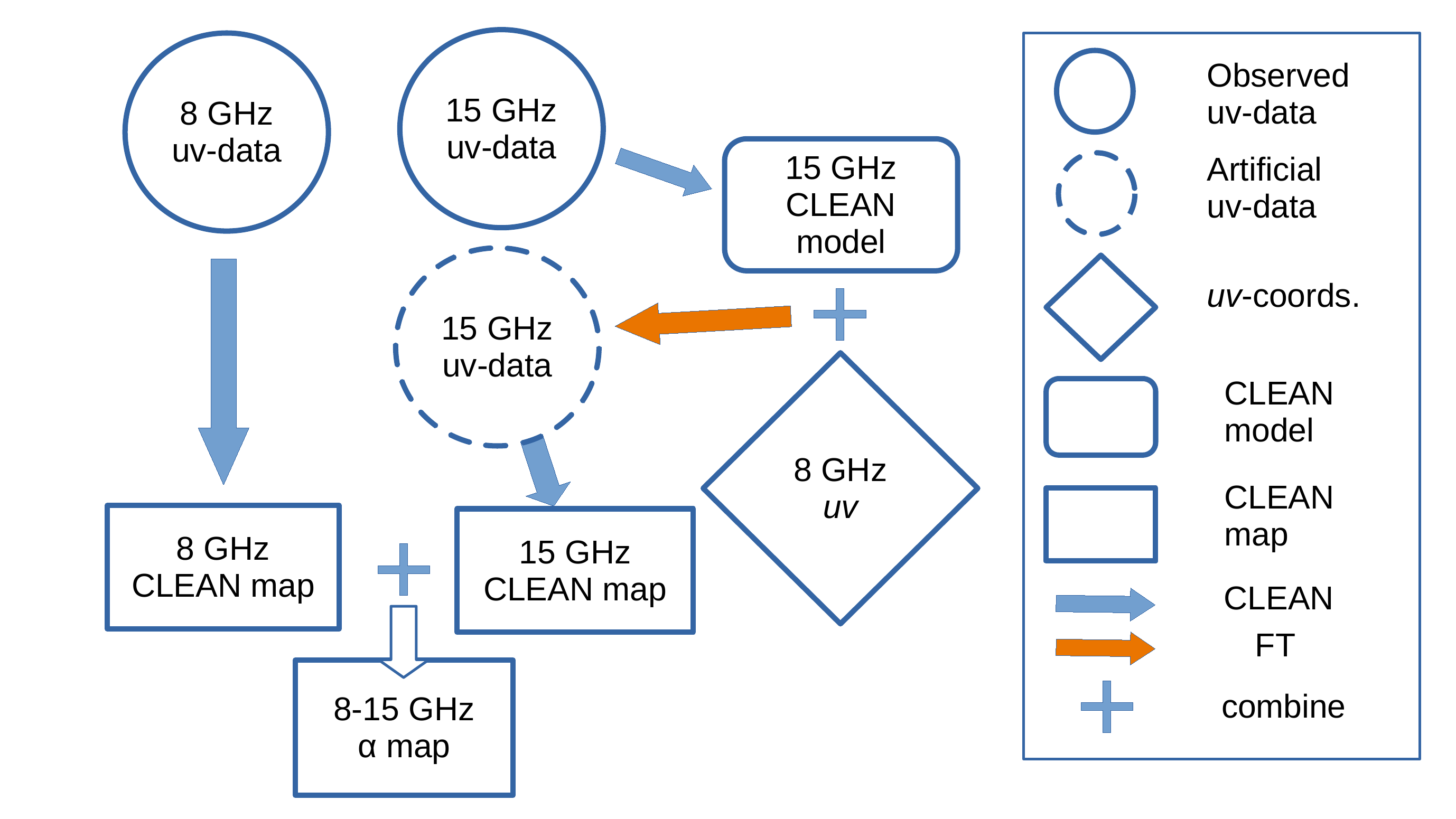}
\caption{Block scheme representing the procedure of the spectral index bias correction \autoref{sec:recreating}.}
\label{fig:scheme_recreation}
\end{figure*}

\begin{figure*}
\centering
\includegraphics[width=0.8\linewidth]{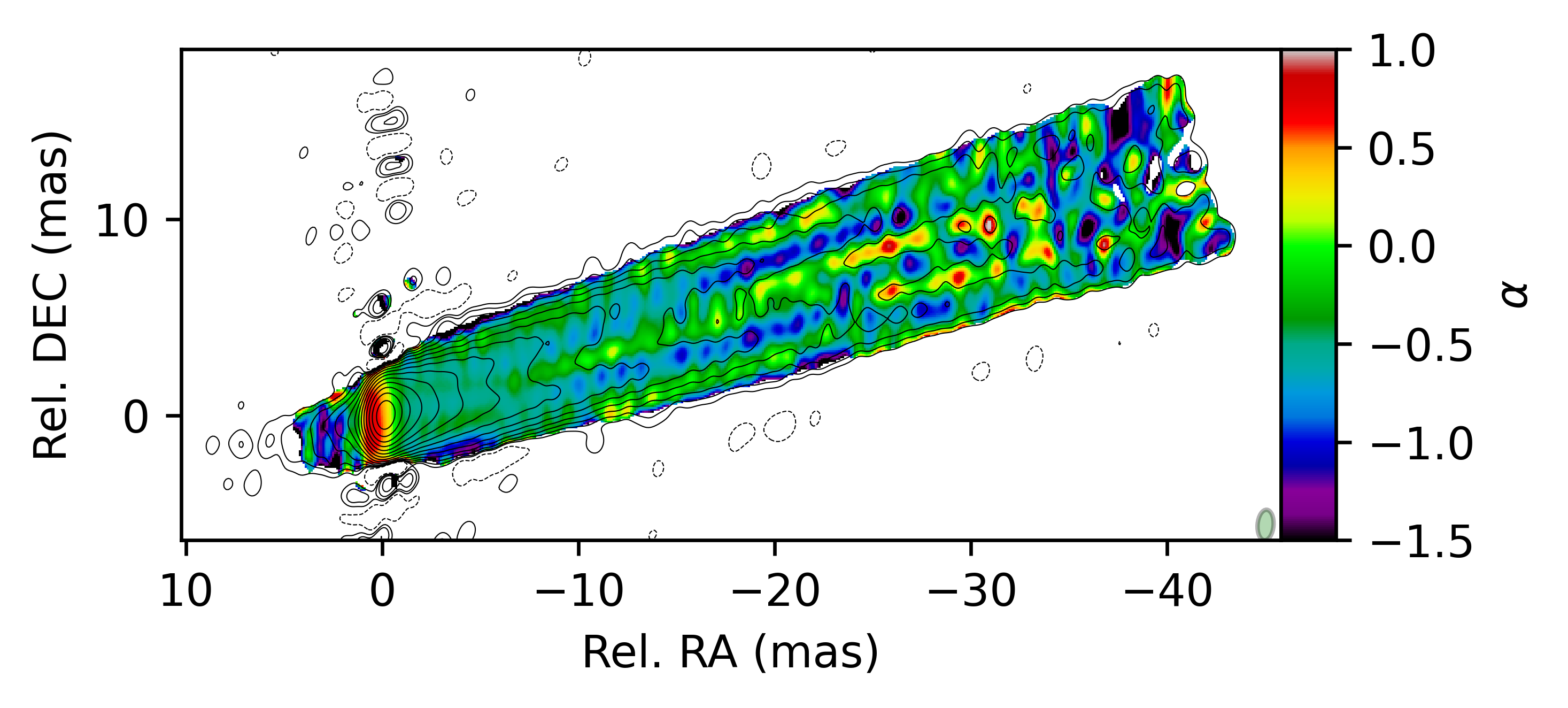}
\includegraphics[width=0.8\linewidth]{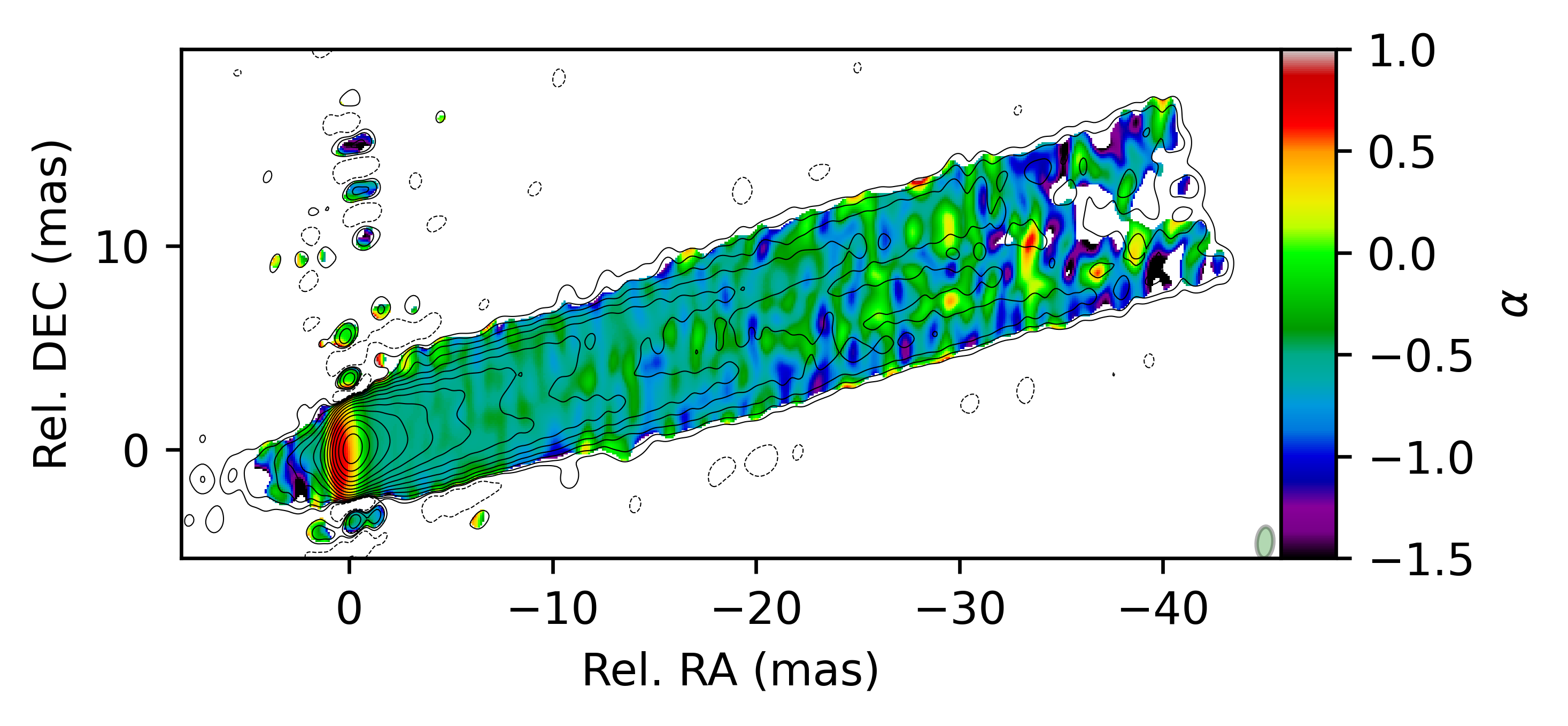}
\includegraphics[width=0.8\linewidth]{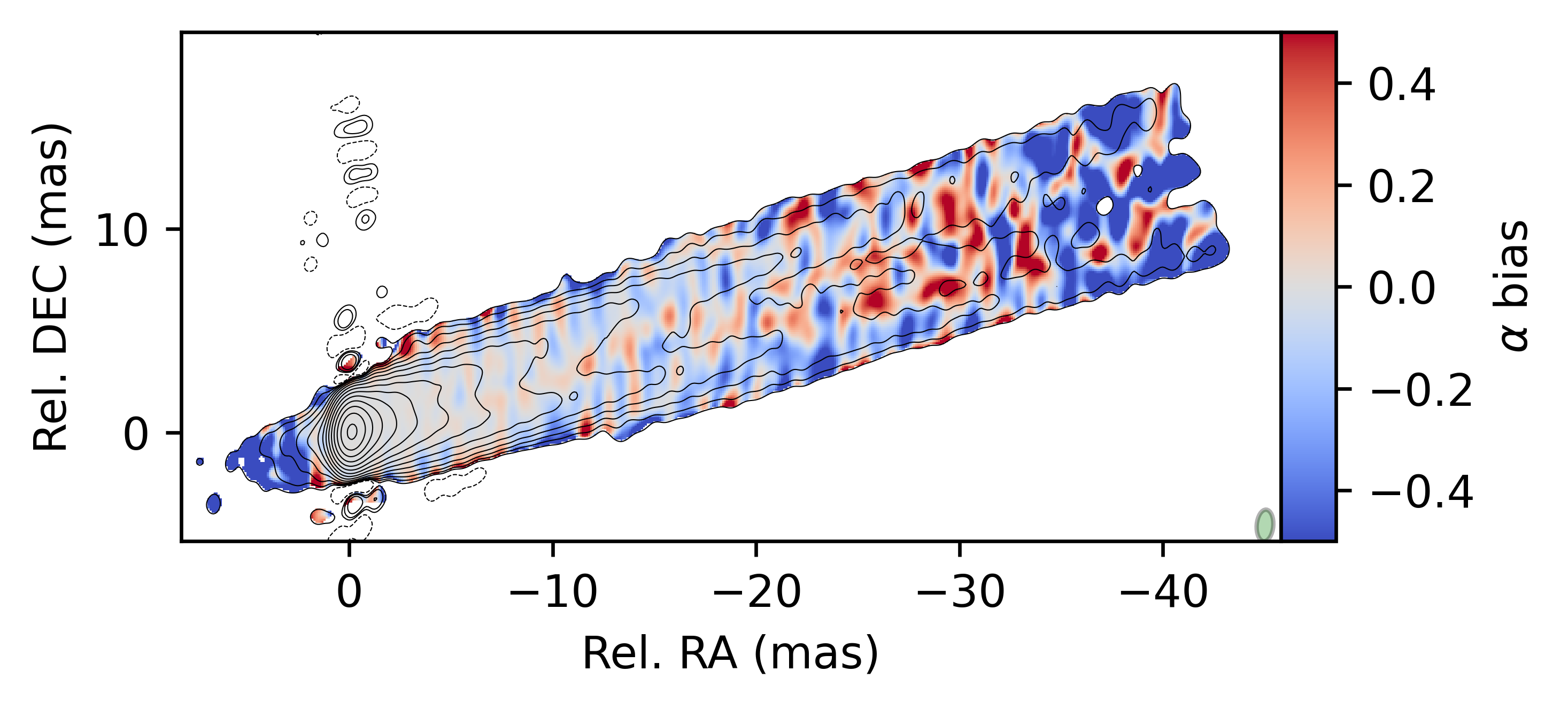}
\caption{Original synthetic spectral index image (\textit{top}), its corrected version (\textit{middle}) with the corresponding residual bias estimate (\textit{bottom}) for BK145 data set and ``2 ridges'' model with $R_{1{\rm pc}} = 0.12$ pc and uniform weighting beam. The correction was obtained by re-imaging 15\,GHz data set on the ($u$,$v$)-coverage of 8\,GHz data set (see \autoref{sec:recreating} for details).}
\label{fig:method2}
\end{figure*}

In \autoref{sec:origin} we found that the artificial stripes of the spectral flattening are due to the systematics in the low frequency Stokes $I$ image. Thus, another way to account the spectral index bias is to re-created the error pattern in a high frequency image. Assuming that Stokes $I$ image bias is relatively small it will cancel out in the expression for the spectral index. To check this on BK145 experiment we created 8 GHz synthetic ($u$,$v$)-data set by substituting the original visibilities with the Fourier Transform of 15 GHz CLEAN model and adding noise to the model visibilities. Then we imaged the obtained data set with the same parameters as the 8 GHz data set and obtained the CLEAN image at 15 GHz with 8 GHz resolution. Schematic representation of the procedure is presented in \autoref{fig:scheme_recreation}. This procedure could be considered as an alternative to the convolving the original 15 GHz CLEAN model with 8 GHz beam. Being an essentially an interpolation, this procedure can be done reliably only if an array have a good ($u$,$v$)-plane coverage (cf. Space VLBI), the frequency range is not large and visibility SNR ratios are comparable (cf. scatter-broadened sources at long wavelengths).
%Strictly speaking, such procedure should be done without $uv$-clipping of the data sets at the maximal $uv$-distance of the low-frequency data set. Indeed, in that case high frequency CLEAN model visibilities at low frequency $uv$-domain will be less affected by CLEAN errors.
% remove sentence?
The resulting Stokes $I$ image showed the same bias as the 8 GHz data set. However the corresponding spectral index map revealed no significant flattening systematics (middle \autoref{fig:method2}). The artificial steepening of the spectral index in the outer jet region and counter jet is still visible, although with a smaller amplitude.
CLEAN is known to interpolate well in ($u$,$v$)-plane and its errors are dominated by ($u$,$v$)-plane extrapolation, not holes in the sampling pattern or lack of zero spacings \citep{briggs}. The suggested procedure employes this and, thus, is a promising method for obtaining an unbiased spectral index maps. It is used for obtaining the unbiased spectral index image of the real BK145 data set in a companion paper (Nikonov et al., in prep.)

\subsection{Avoiding spectral steepening bias}
\label{sec:correct_steepening}
\begin{figure*}
\centering
\includegraphics[width=0.73\linewidth]{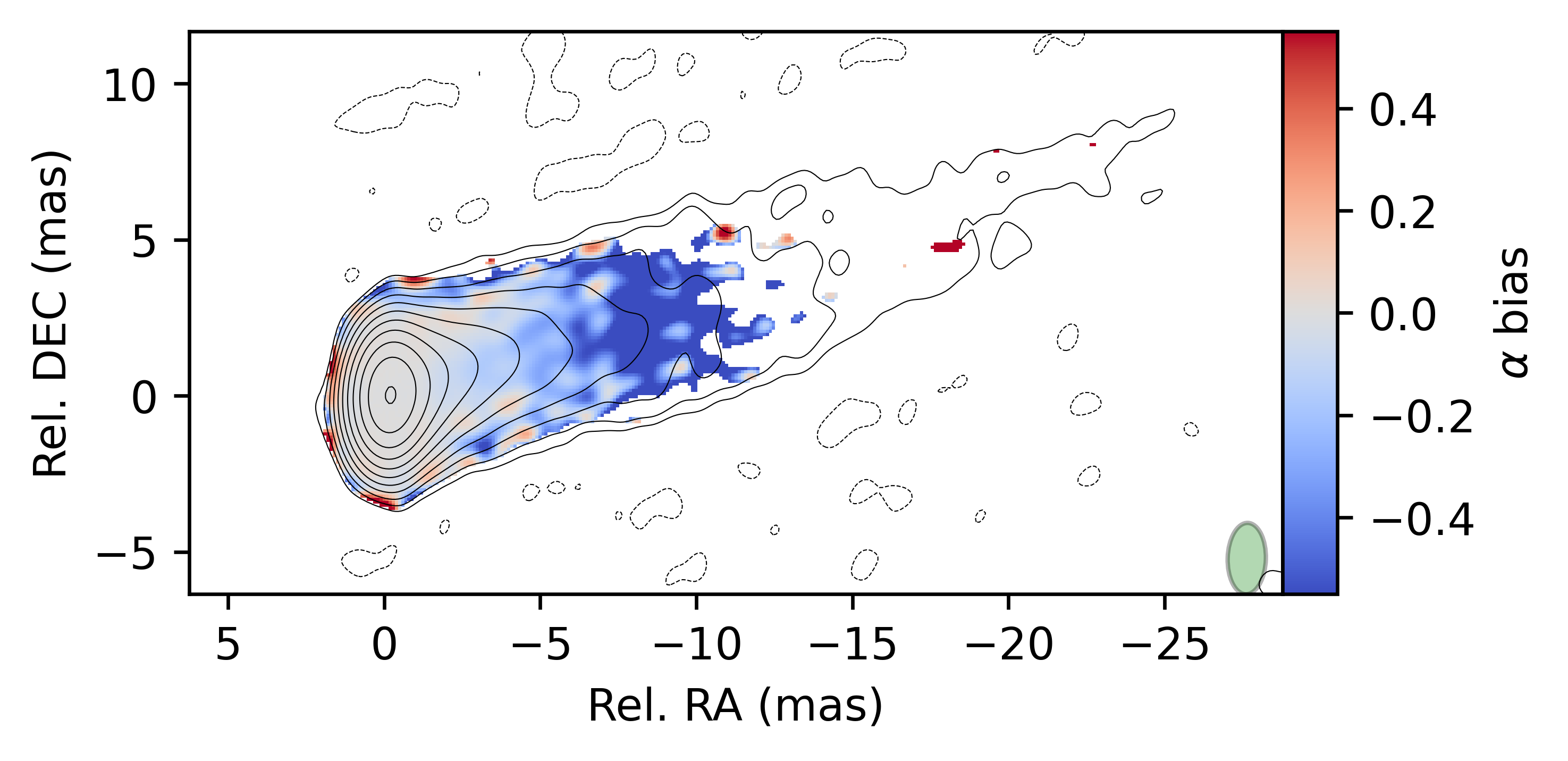}
\includegraphics[width=0.73\linewidth]{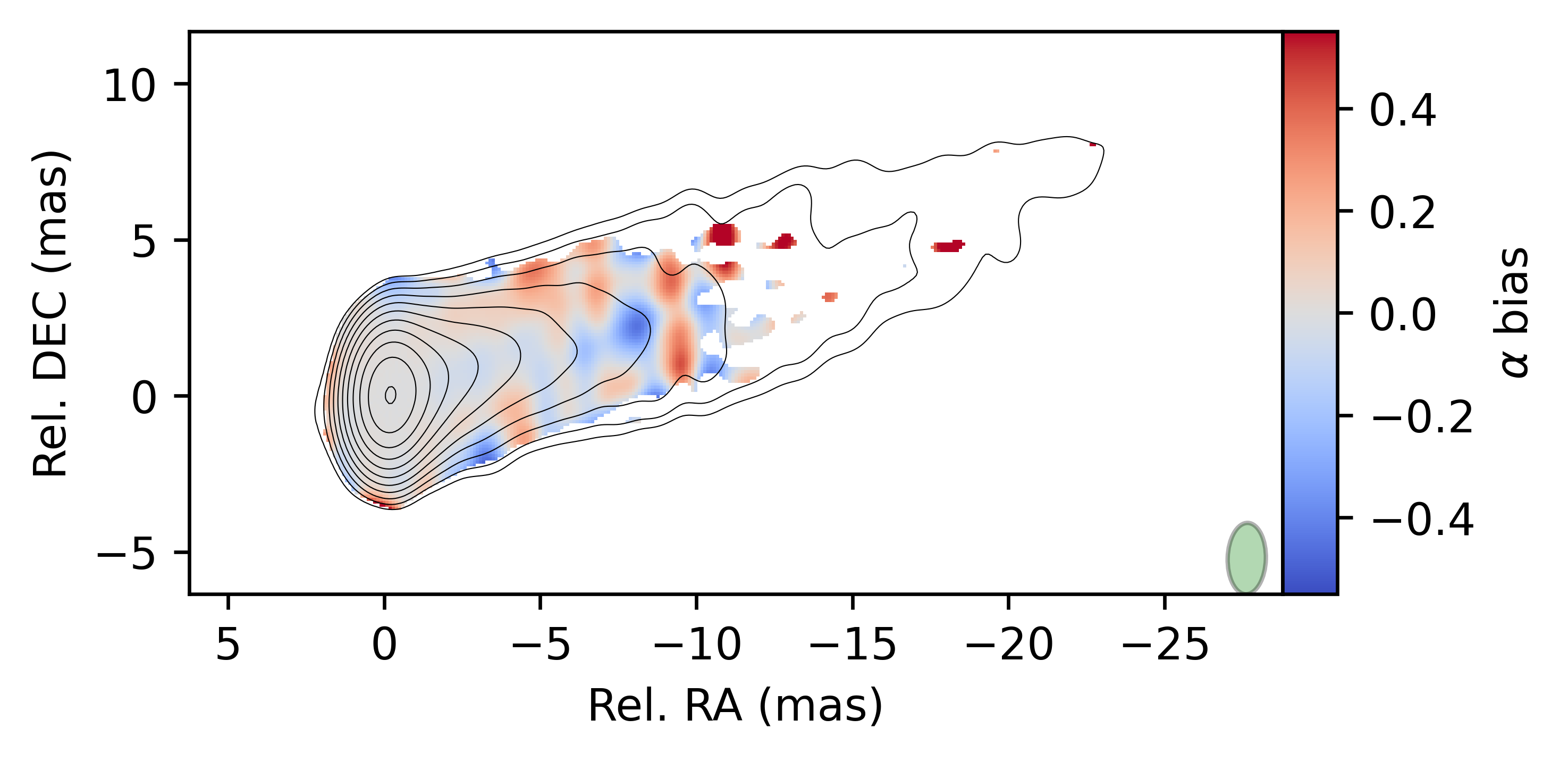}
\caption{Bias of the spectral index for conventional CLEAN (\textit{Top}) and deep CLEAN (\textit{Bottom}) for MOJAVE 8 and 15 GHz and ``KH'' model brightness. The lowest positive contour and the masking of the bias are the same for comparison and were determined by the conventional CLEAN results.}
\label{fig:deep}
\end{figure*}

The correction for the spectral steepening bias naturally follows from its nature, i.e. the residuals of the high frequency CLEAN image. The simplest way is to CLEAN deep into the noise to move all flux to the CLEAN components \citep{briggs,1999ASPC..180..301F}. In \autoref{fig:deep} the spectral index bias maps are presented for MOJAVE 8 and 15 GHz simulated data set for both conventional CLEANing up to the noise \textit{rms}, estimated at 1 arcsecond away from the phase center and deeper CLEANing well below the \textit{rms} level. ``KH'' model brightness \autoref{fig:JetModels} was employed as a ``true'' model for bias estimation. Uniform weighting was used in both cases and stacking was performed to focus on the systematics. The extended region of negative bias in the outer jet region disappeared. Instead, negative and positive fluctuations with the amplitude less than a random error and a typical scale of a beam size are seen. The other possible way to correct for the bias is to scale the residuals to the ratio between the area of the common convolving beam and the high frequency dirty beam. However, this only partially reduces the bias because the residual flux stays more concentrated due to the smaller effective size of the high frequency dirty beam.

The procedure of the spectral steepening bias correction by deep CLEAN could be complicated by the dynamical range effects in a real data, e.g. the residual calibration errors or deconvolution errors encountered in our simulations. This could prevent CLEAN to go below the $rms$ level. Thus, we checked the procedure on a real data sets. We made use four frequency (8.1, 8.4, 12.1 and 15.4 GHz) MOJAVE multifrequency data\footnote{\url{https://www.cv.nrao.edu/MOJAVE/allsources.html}} and choose several sources with different morphology, showing sings of the spectral steepening. Corresponding Stokes $I$ visibilities at each band were imaged in two ways. The first one -- the original used by the MOJAVE team. The second one -- deep CLEAN up to 0.1 of the $rms$ level. Then the spectral index was estimated according to \cite{MOJAVE_XI} for both sets of the obtained multifrequency images. The difference between the conventional and deep CLEAN spectral images for several sources is presented in \autoref{fig:deep_clean_diff}. Here the negative value (blue colour) represents steeper spectral index for conventional CLEAN. Interesting that shallow CLEAN could not only steepen the spectra in the extended jet regions, but also flatten it at the position of well isolated components (e.g. in 1652$+$398).

\subsection{Avoiding CLEAN biases}
\label{sec:avoidCLEANbiases}

In previous sections we showed how spectral index biases could be compensated using only data at hand. Here we discuss how the CLEAN images could be constructed in a least biased way.

As mentioned in \autoref{sec:origin}, the images of extended objects deconvolved with a point-source CLEAN tend to show systematic corrugations that can be confused for real structure \citep{Cornwell_1983,1984iimp.conf..255S,1984A&A...137..159S,1999ASPC..180..151C,briggs}.

There are several approaches to CLEAN to improve
this. The algorithm proposed by \cite{Cornwell_1983} consists in adding some smoothing constrains that result in a modified dirty beam. Smoothness reduces the spurious flux at the edge or outside of the sampled ($u$,$v$)-points. The resulting images demonstrate significant improvement with a slight bias of the total and peak flux due to the constrains. However, this solution can be shown to have undesirable properties \citep{10.1093/mnras/220.4.971}.
Another approach is to abandon the assumption that each peak in the residuals is the location of a single component in the object, which is equivalent to the prior assumption for obtaining the narrowest possible features in the image \citep[SDI CLEAN][]{1984A&A...137..159S}.

The performance for extended objects can also be improved using, e.g. a multi-scale approach \citep{2008ISTSP...2..793C}.
In such methods the object is represented of various different scale sizes. This allows for better separation between source image features, noise and sidelobes. For example, multi-scale (MS-CLEAN) removes large-scale structure before finer details \citep{2008ISTSP...2..793C,Rich_2008}. However, simulations of \cite{2008ISTSP...2..793C} show that MS-CLEAN is also liable to fringing artefacts but at a lower level.

Another recent multi-scale approach is wavelets. \cite{Muller_2023} employed a wavelet dictionary consisting of differences of Gaussians of Bessel functions. The method makes a step forward a traditional multi-scale approach \citep[e.g. AsP-CLEAN][]{2004A&A...426..747B} by selecting the scales based on the Fourier domain, thus  suppresses the power in Fourier coefficients on longer than the longest baseline, while keeping the fit to the measured visibilities. Simulations with a several typical cases showed that this approach to be a superior over traditional CLEAN and MS-CLEAN in both over-resolution and recovering the extended structure without introducing a fringing.

Finally, we present a short summary of the various biases encountered in total intensity and spectral index images obtained with CLEAN algorithm and ways of their mitigation in \autoref{tab:short_summary}.

\begin{table*}
    \centering
    \caption{CLEAN-based total intensity and spectral image artefacts and their mitigation.}
    \label{tab:short_summary}
    \begin{tabular}{lllll}
        \hline
         Image & Artefacts & Origin & Mitigation \\
        \hline
        Stokes I & Ridges along the jet & \makecell[cl]{CLEAN errors at the outskirts \\ of the ($u$,$v$)-plane} & Special CLEAN methods (Sec.~\ref{sec:avoidCLEANbiases})\\
        $\alpha$-image & Flattening/steepening stripes along a jet & \makecell[cl]{CLEAN errors (as above) of \\the low frequency map} & \makecell[cl]{Bias correction (Sec.~\ref{sec:recreating}) \\ or special CLEAN methods (Sec.~\ref{sec:avoidCLEANbiases})} \\
        $\alpha$-image & Steepening in the extended regions & High-frequency map residuals & Deep CLEAN (Sec.~\ref{sec:correct_steepening}) \\
        \hline
    \end{tabular}
\end{table*}

\section{Astrophysical implications}
\label{sec:implications}

We demonstrated that CLEAN produces specific spurious details in the VLBI total intensity and spectral index images that can be interpreted physically. Thus, we suggest that any structure in the spectral index maps should be considered as robust only when obtained using different imaging techniques or with a simulations addressing possible biases. 

% Note we are talking about the central part of the jet, not KH-modes
We showed that the emission at the central part (spine) of the M87 jet at parsec scales could be consistent with a constant optically thin spectral index. 
% We showed that the jet emission in M87 at parsec scales could be consistent with a constant optically thin spectral index.
The absence of the central spectral flattening implies that the optical depth in the central jet region is low and jet emission is optically thin.
Possible geometrically thin optically thick ``core'' in the centre of the jet should contribute negligibly to the observed flux not to imprint itself in the flattened spectral index \citep{Zakamska_2008}. 

It also rules out high values of the low energy cutoff  $\gamma_{\rm min}$ in the emitting particles energy distribution for the central ``core'' with a high magnetic field or quasi-monoenergetic distribution. 
Although such models could still be consistent with the constant spectral index distribution if the emissivity is distributed non-uniformly across the jet radii and the heated particles are concentrated close to the jet edges \citep[e.g. through magnetic reconnection triggered by KH, ][]{2021ApJ...907L..44S}, thus avoiding the central regions with high magnetic field \citep{Beskin09,Kom09,Lyu09}. This also holds for the particles heating prescription of \cite{2021arXiv211102518F}, where the central part of the jet with a higher magnetization displays flatter spectral index.

At the same time our results are also consistent with velocity stratified jet models where the flux depression in the fast spine is due to the Doppler de-boosting, which is achromatic. Recent analysis of the implicitly measured jet speeds \citep{2022MNRAS.509.1899N} implies that jets can accelerate up to relatively high Lorentz factors $\Gamma \sim 100$ which is consistent with observations of edge brightening in M87 and 3C 84 \citep{2016ApJ...833...56A,2018ApJ...855..128W,RA_3C84}. However, as Doppler effect shifts the emitting frequency, the absence of the observed optically thick emission in M87 jet spine implies that the lower frequency cutoff $\nu_{\rm SSA}$ due to synchrotron self-absorption should be less than the emitted frequency $\nu_{\rm obs}/D_{\rm spine}$, where $\nu_{\rm obs}$ - the observed frequency and $D_{\rm spine}$ - the spine Doppler factor. This is easily achieved in radiogalaxies, but could be violated for blazars with $D_{\rm spine} >\!\!> 1$. Interesting, that \cite{2021A&A...654A..27B} observed flattening spectra at the jet spine of 3C273 with high-resolution \textit{RadioAstron} observations at 1.6 and 4.8 GHz. However, the jet is edge brightened at 1.6 GHz and core brightened at 4.8 GHz. This rules out the velocity stratification as a sole reason of the effect \citep{2021A&A...654A..27B}. As discussed in \autoref{sec:intro}, spectral flattening due to the lower energy cutoff $\gamma_{\rm min}$ also depends on the Doppler factor. Indeed, emission at a given frequency $\nu_{\rm obs}$ is dominated by electrons with Lorenz factors $\gamma_{\rm rad} \approx \sqrt{\nu_{\rm obs} \Gamma / (D \nu_{B})}$, where $\nu_{B}$ - Larmor frequency for magnetic field and we assumed the dominant toroidal component. That could reconcile the spectral flattening in the jet centre and different brightness profiles at the both frequencies observed by \cite{2021A&A...654A..27B} with the fast spine scenario.

% Our simulated spectral images also reveal some discrepancies with the observed maps. First, as can be seen from \autoref{fig:alpha_2ridges_bk145_xbeam}, the flattest spectral index in the core region $\alpha_{\rm core,sim} \approx 0.75$ is slightly more inverted than the observed $\alpha_{\rm core,obs} \approx 0.3-0.4$ \citep{nikonov}. The mean core spectral index between 15 and 8 GHz for MOJAVE sample is $\alpha \approx 0.2$ \cite{MOJAVE_XI}, while the median core spectral index between 8 and 2 GHz for a sample of 370 AGNs is $\alpha = 0.3$.
% WE ARE NOT MODELLING DATA!
% However, the uncertainty of the spectral index in the core region due to image alignment error is relatively high for straight jets without prominent knots \citep{MOJAVE_XI}. Note also that our simulated images alignment is precise by construction. If the difference is real, it could imply bulk acceleration in the core region \citep{2019MNRAS.489L..58Z}. Another possible reason is the low energy cutoff $\gamma_{\rm min}$. As discussed in \autoref{sec:intro}, in that case the self-absorption spectral index is less inverted: $\alpha_{\rm SSA} = 2.0$ instead of $2.5$. Further down the jet $\alpha$ should steepen to optically thin $\alpha_{\rm cutoff} = 1/3$ and then to its optical thin value $\alpha = -(s-1)/2$. 

Another common bias encountered in our simulations is evident at the low SNR jet regions: the simulated $\alpha$ is typically steeper than the true value. 
\cite{pushkarev_etal12} analysed dual frequency (2 and 8 GHz) VLBA and geodesic VLBI observations of 370 AGNs and found that the spectral index typically decrease along the jet with the median value $-0.05$ mas$^{-1}$. However they did not performed $uv$-matching, making this value an upper limit. 
Steepening of the spectral index was also found for 191 sources from MOJAVE sample observed at four frequencies from 8 to 15 GHz \citep{MOJAVE_XI}. The typical steepening between the edge of the convolved core and the median jet spectral index is $0.52\pm0.03$ for quasars and $0.39\pm0.06$ for BL Lacs objects. It was interpreted as radiative losses or high-energy cutoff in the electron spectrum, depending on the jet geometry. Although \cite{MOJAVE_XI} employed simulations with CLEAN models and showed that systematics could not explain the observed steepening, our simulations with a more realistic brightness models demonstrate that the observed spectral index images of M87 are consistent with a constant $\alpha$. If it is indeed the case it could imply continuous re-acceleration of the emitting particles along a jet, e.g. via magnetic reconnection \citep{2021ApJ...907L..44S}. Interesting, that \cite{MOJAVE_XI} also found that spectral index flattens at the position of the jet components. The mean difference with the median jet spectral index was estimated as $0.2\pm0.02$. We found that the steepening bias depends on the SNR, \autoref{sec:steepening}. Thus, the observed spectral flattening at component location could be actually due to the spectral steepening in the low SNR jet regions around components (e.g. 1652$+$398 in \autoref{fig:deep_clean_diff}). This, in turns, could imply an absence of the ongoing particle acceleration in components that flattens the spectrum \citep{MOJAVE_XI}. In other words, some components could be moving blobs instead of travelling shocks. However, our models given in \autoref{sec:Model} describe smooth and continuous jet without emission enhancement at a positions of jet components. Thus, further investigation is necessary.

\section{Summary}
\label{sec:conclusions}

We employed a series of a dual frequency simulations with the realistic jet brightness models to explore the possible biases in the spectral index VLBI maps. Comparing the results with the published high-fidelity images we found that the spectral maps of M87 radio jet are heavily affected by the imaging systematics. It both flattens the spectra in a series of stripes elongated along a jet and steepens it in a low brightness extended jet region. For the first kind of the systematics the origin is the CLEAN errors at the edges or just outside of the sampled ($u$,$v$)-points in the low-frequency image. The steepening spectral bias is due to the uncleaned residual flux in the high-frequency image. We propose methods for compensating the systematics using only the observed data and demonstrate their efficiency both on simulated and real data sets. We also show that inner ridge line in M87 radio jet could be a CLEAN imaging artefact.
Our results suggests that the obtained so far spectral index images of M87 parsec jet could be consistent with a constant optically thin intrinsic value. This could imply the continuous particle acceleration process acting along a jet, e.g. magnetic reconnection.
The detailed analysis of the Global VLBI dual-frequency spectral index image of M87 jet, accounting for the discovered biases is presented in a companion paper (Nikonov et al., in prep.).

\section*{Acknowledgements}
We thank the anonymous referee for helpful comments and suggestions that significantly improved the presentation of the results. We thank Daewon Kim for valuable comments.
We thank Yuri Y. Kovalev for bringing our interest to the subject.
This study has been supported by the Russian Science Foundation, project\footnote{Information about the project: \url{https://rscf.ru/en/project/20-72-10078/}} 20-72-10078.
This research has made use of data from the MOJAVE database that is maintained by the MOJAVE team \citep{Lister-19}. The MOJAVE program is supported under NASA-Fermi grant 80NSSC19K1579.
This research made use of data obtained from Data ARchives and Transmission System (DARTS), provided by Center for Science-satellite Operation and Data Archive (C-SODA) at ISAS/JAXA \citep{darts1,darts2}.
The National Radio Astronomy Observatory is a facility of the National Science Foundation operated under cooperative agreement by Associated Universities, Inc.

This research made use of \textit{Astropy}, a community-developed core Python package for Astronomy \citep{2013A&A...558A..33A}, \textit{Numpy} \citep{numpy}, \textit{Scipy} \citep{scipy}, \textit{FINUFFT} \citep{2018arXiv180806736B}. \textit{Matplotlib} Python package \citep{Hunter:2007} was used for generating all plots in this paper.

\section*{Data availability}

This research made use of the data from the MOJAVE database\footnote{\url{https://www.cv.nrao.edu/MOJAVE}} which is maintained by the MOJAVE team \citep{MOJAVE_XV}.
This research made use of NASA's Astrophysics Data System.
The visibility data that was used as a template for ($u$,$v$)-coverage and noise estimation is available in \cite{2020A&A...637L...6K}, Nikonov et al. (in prep.) and DARTS archive\footnote{\url{http://www.darts.isas.jaxa.jp}}. 
The data generated in this research will be shared on reasonable request to the
corresponding author.

\bibliographystyle{mnras}
\bibliography{nee}

\appendix

\section{Model details}
\label{a:model_physics}

To model the observed transverse stratification, we modulated the transverse profile of the heated particles in a several ways.
The uniform transverse profile, i.e. no transverse modulation of $N \propto z^{-n}$ dependence, was set as a base model (hereinafter -- ``BK'' model, from ``Blandford-K\"{o}nigl'').
To model the double (edge-brightened) and triple (edge-brightened with a central spine) emission profiles, we assumed that the emitting particle density at a given distance from the jet origin $z$ is modulated as $\propto \exp{(-(r_{\rm frac} - r_{\rm c})^2/\sigma_{r}^2)}$, where $r_{\rm frac}$ - fraction of the jet radii at a given point and $\sigma_{r} = 0.025$. 
For a double emission profile (edge-brightened) we set $r_{\rm c}=0.9$. For a triple humped emission profile we use $r_{\rm c}=0.0$ and $r_{\rm c}=0.9$.  We prefer setting emitting particles at certain radii (0.0 and 0.9 fraction of a jet radius) in contrast with setting them at certain magnetic field line \citep[e.g.][]{2009ApJ...697.1164B,2019ApJ...877...19O} due to the following reasons. The central hump ($r_{\rm c}=0.0$) in emitting particles number density is associated with a central core with an almost constant radius for a core limiting magnetic field surface. Although an accelerating jet needs magnetic field bunching \citep{KBVK07, TMN09, Nakamura+18}, the latter affects mostly the central part of a jet, and the edges only to a small degree. We have checked on analytical MHD models \citep{Lyu09, BCKN-17} that the inner field lines $\Psi=0.1\Psi_0$ and $\Psi=0.5\Psi_0$ have more collimated shape than the jet boundary. But for $\Psi=0.9\Psi_0$ the shapes do not differ drastically, so we can use a constant fraction of a jet radius as a proxy for the constant magnetic surface if it is close enough to a jet boundary. Another argument is connected with a possible de-boosting effect and connection of the observed emission with the plasma bulk velocity. We discussed in \citet{2022MNRAS.509.1899N} that the surface with maximum Lorentz factor has a shape similar to the one of a jet boundary. Thus, a choice of a certain fraction of a jet radius to set a hump emitting particle spacial distribution is preferable in this case.
 
The three ridge heating in our model mimics two different possible sources of a jet emission.
The central part of a jet is known to have a core-like structure with local maximums in particle number density and magnetic field amplitude on the scales of a few light cylinder radii \citep{Beskin09,Kom09,Lyu09}. 
Even for a small fraction of heated particles the region may be locally prominent in emission.
Thus, the central ridge $r_{\rm c}=0$ reflects this MHD jet feature. 
The edges heating, on the other hand, may be due to plasma effective heating at the jet boundary --- in the layer between a jet and a collimating ambient medium / wind. \citet{McKinney06} and later \citet{Chatterjee19} showed developed pinch instabilities along a jet, which may lead to emission due to both plasma heating on shocks and local variations in optical depth. 
In addition, plasma instabilities at the jet boundary create conditions for a magnetic reconnection which also heats plasma and leads to emission \citep[see e.g.,][]{SS14, Cerutti15,2021ApJ...923L...5P}. Another possible source of edge brightening could be a flow velocity
stratification. Analytical \citep{Kom09, BCKN-17} and numerical \citep{Chatterjee19} modeling demonstrates that centrifugal MHD acceleration works inside
a fast jet part (spine) due to constant angular velocity of field lines.
The outer part of a jet (sheath) could be slower due to decreasing
angular velocity \citep[e.g., models with a closed electric current][]{KBVK07, BCKN-17},
interaction with slower wind or ambient medium \citep{Chatterjee19} or both.
In this case edge emission is brighter due to the de-boosting effect.
% We prefer setting emitting particles at certain radii (0.0 and 0.9 fraction of a jet radius) in contrast with setting them at certain magnetic field line \citep[e.g.][]{2009ApJ...697.1164B,2019ApJ...877...19O} due to the following reasons. The central hump ($r_{\rm c}=0.0$) in emitting particles number density is associated with a central core with an almost constant radius for a core limiting magnetic field surface. Although an accelerating jet needs magnetic field bunching \citep{KBVK07, TMN09, Nakamura+18}, it plays role mostly in the central part of a jet. We have checked on analytical MHD \citep{Lyu09, BCKN-17} models that the inner filed lines $\Psi=0.1\Psi_0$ and $\Psi=0.5\Psi_0$ have more collimated shape than the jet boundary. But for $\Psi=0.9\Psi_0$ the shapes does not differ drastically, so we can use a constant fraction of a jet radius as a proxy for the constant magnetic surface if it is close enough to a jet boundary. Another argument connected with a possible de-boosting effect. We discussed in \citet{2022MNRAS.509.1899N} that the shape of a line with maximum Lorentz factor has a close shape to a jet boundary. So, a choice of a certain fraction of a jet radius to set an emitting particle is preferable.

Finally, we employed modulation of the transverse heating profile with two 3D spirals as a model of the Kelvin-Helmholtz-generated heating \citep[hereinafter ``KH'' model;][]{2011ApJ...735...61H}, e.g. through triggered magnetic reconnection \citep{2021ApJ...923L...5P}. We used the spiral radius equals to 0.05 of the local jet radius. In all cases we also considered some background non-thermal particles density with fraction $f_{\rm bg} = 0.01$.

\section{BK145 CLEAN images at 15 and 8 GHz}
\label{a:15ghz}

\begin{figure*}
\centering
\includegraphics[width=0.8\linewidth]{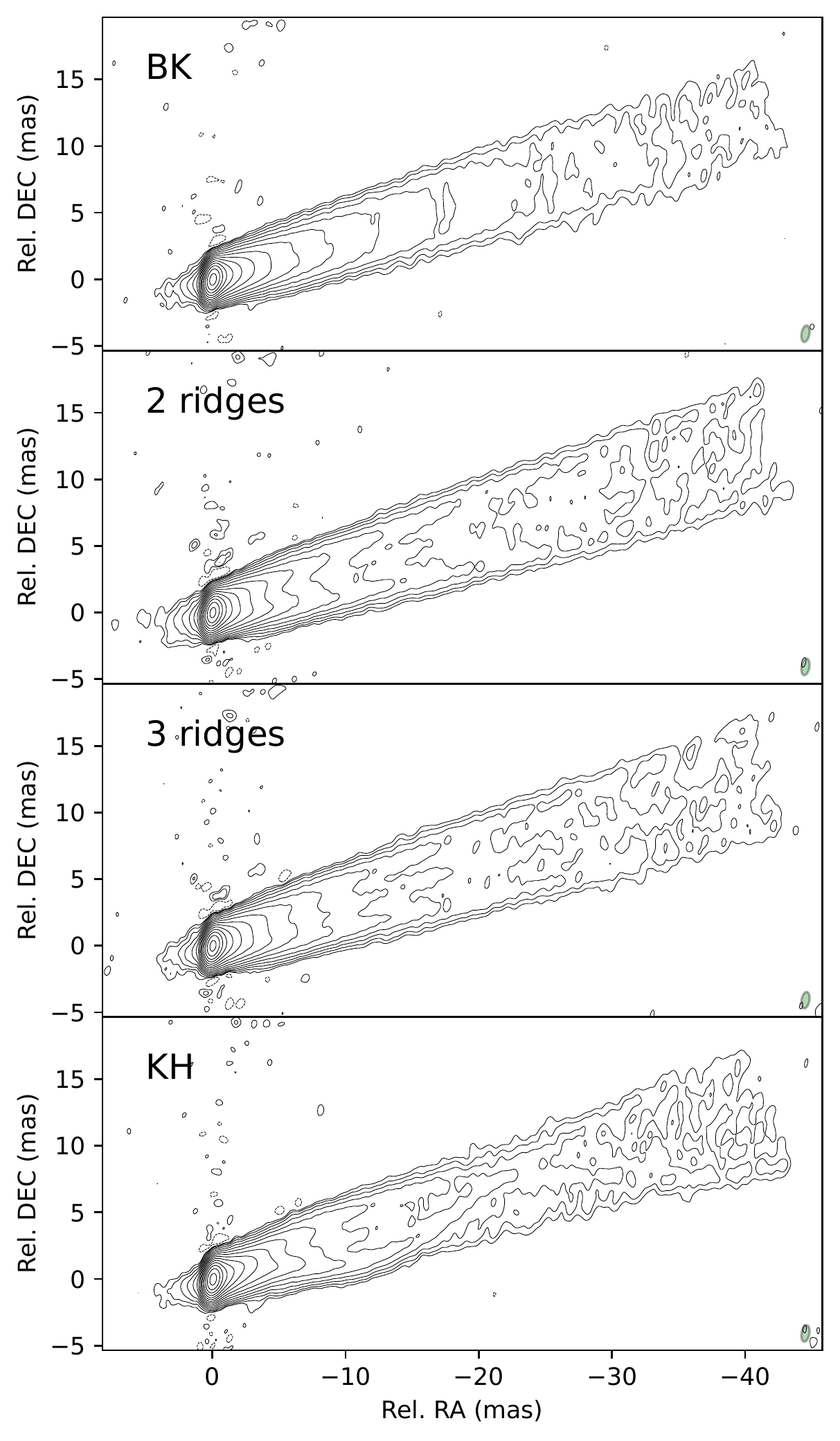}
\caption{The 15 GHz CLEAN images for four jet models with $R_{1{\rm pc}} = 0.12$ pc.}
\label{fig:bk145_15GHz_CLEAN_allmodels}
\end{figure*}

\begin{figure*}
\centering
\includegraphics[width=0.8\linewidth]{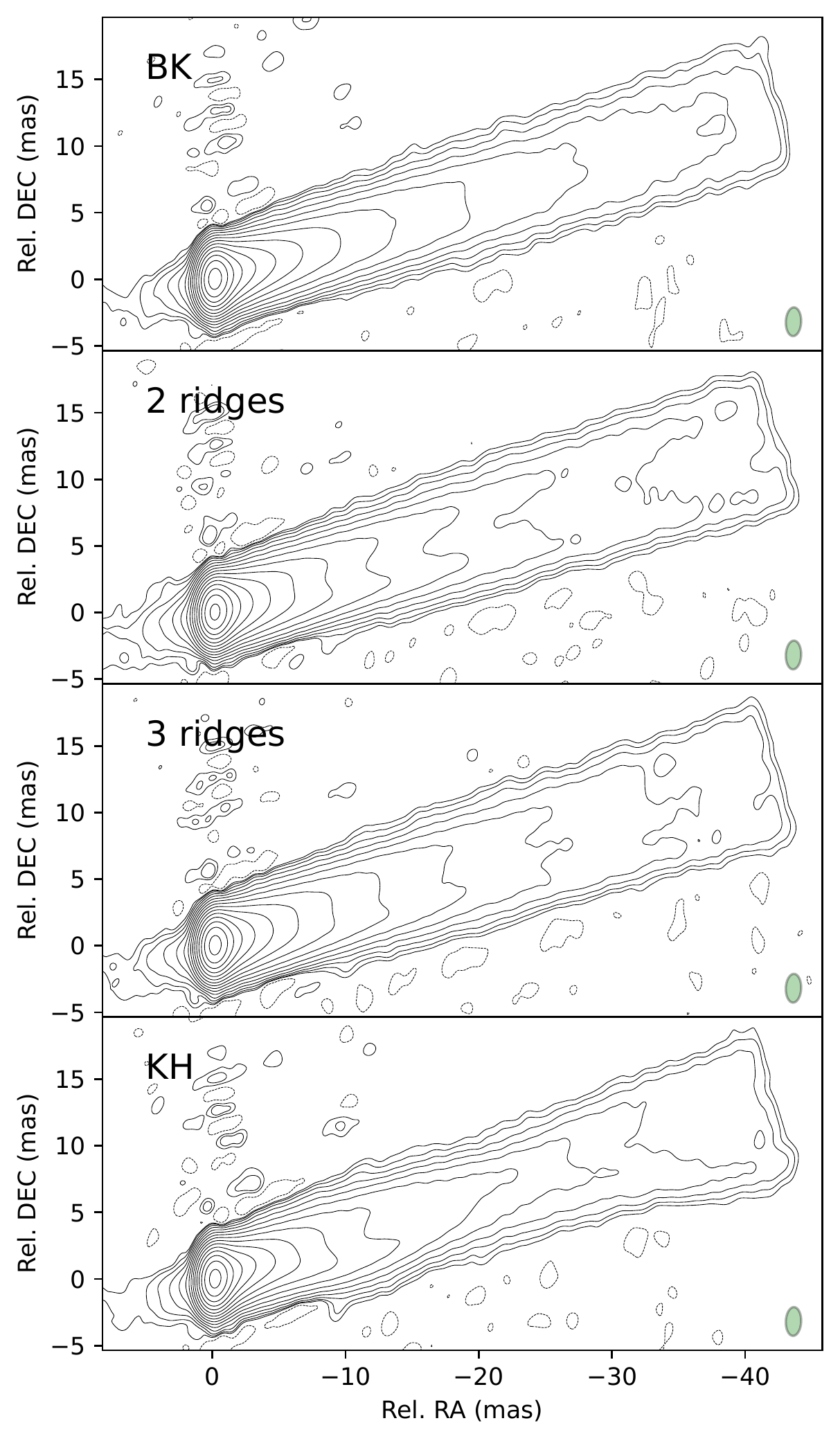}
\caption{The 8 GHz CLEAN images for four jet models with $R_{1{\rm pc}} = 0.12$ pc.}
\label{fig:bk145_8GHz_CLEAN_allmodels}
\end{figure*}

\section{Deep CLEAN spectral index images}
\label{a:deepclean}

\begin{figure*}
    \centering
    \includegraphics[width=0.49\linewidth]{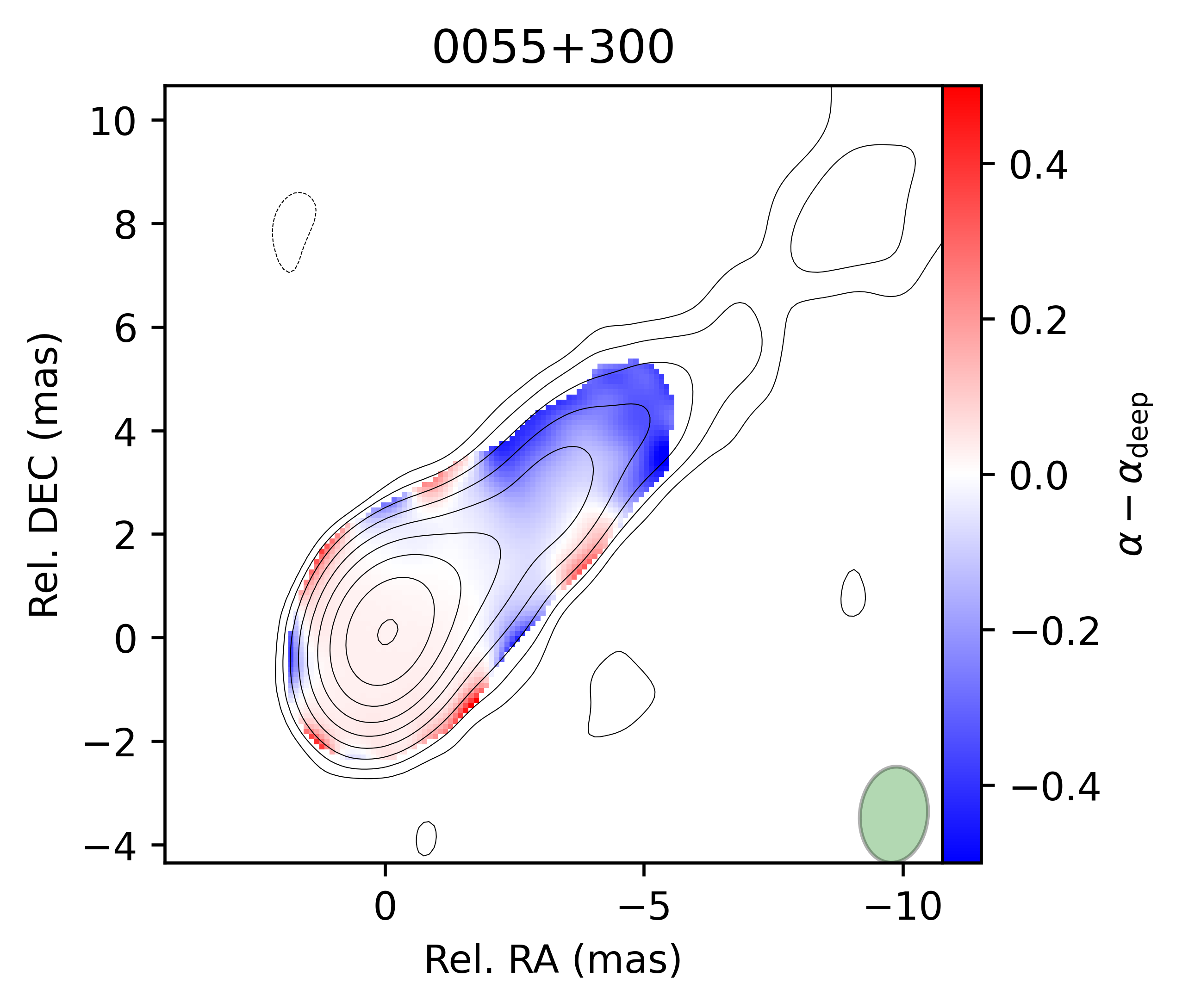}\hspace{0.1cm}
    \includegraphics[width=0.49\linewidth]{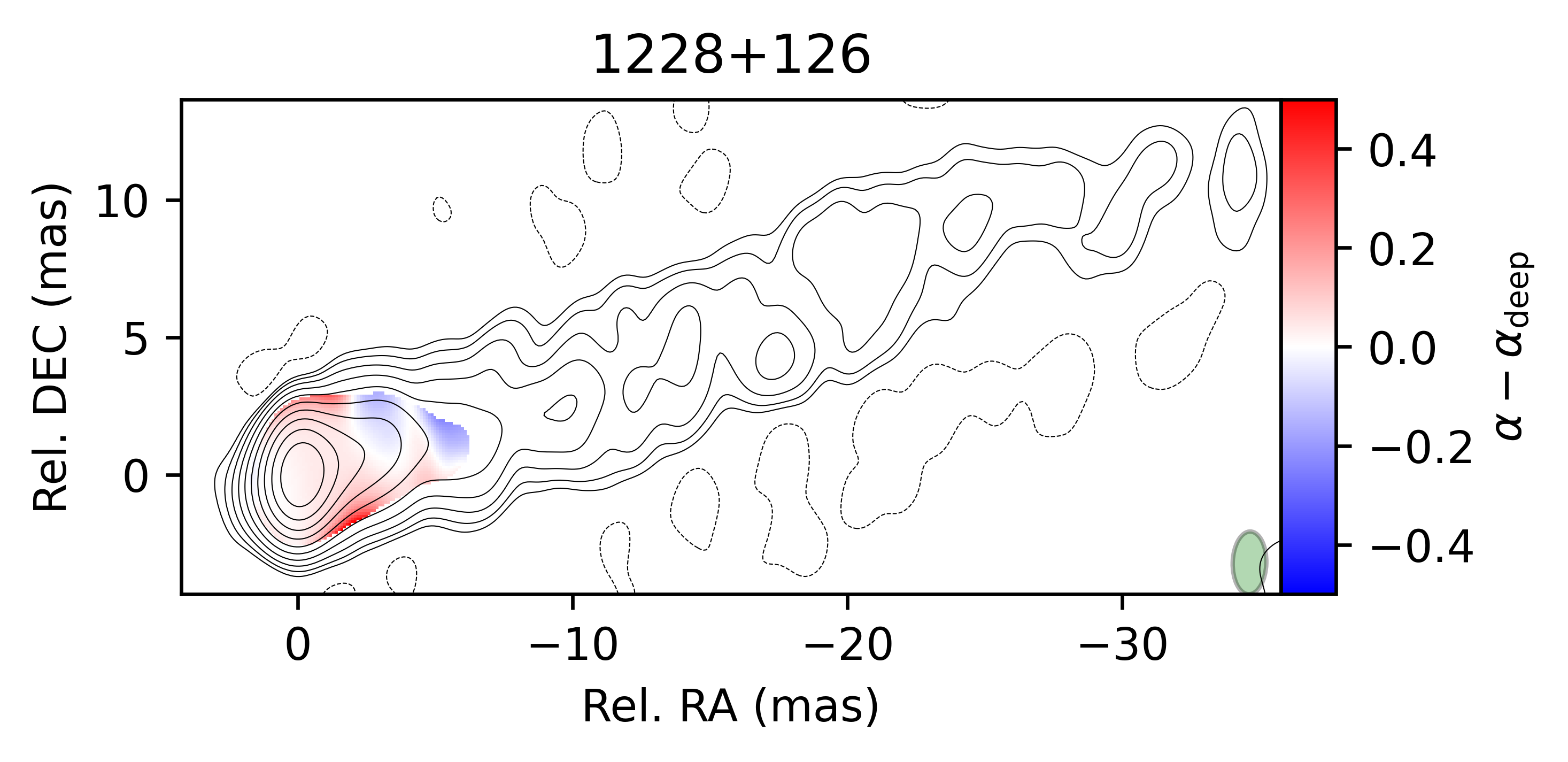}\vspace{0.5cm}
    \includegraphics[width=0.49\linewidth]{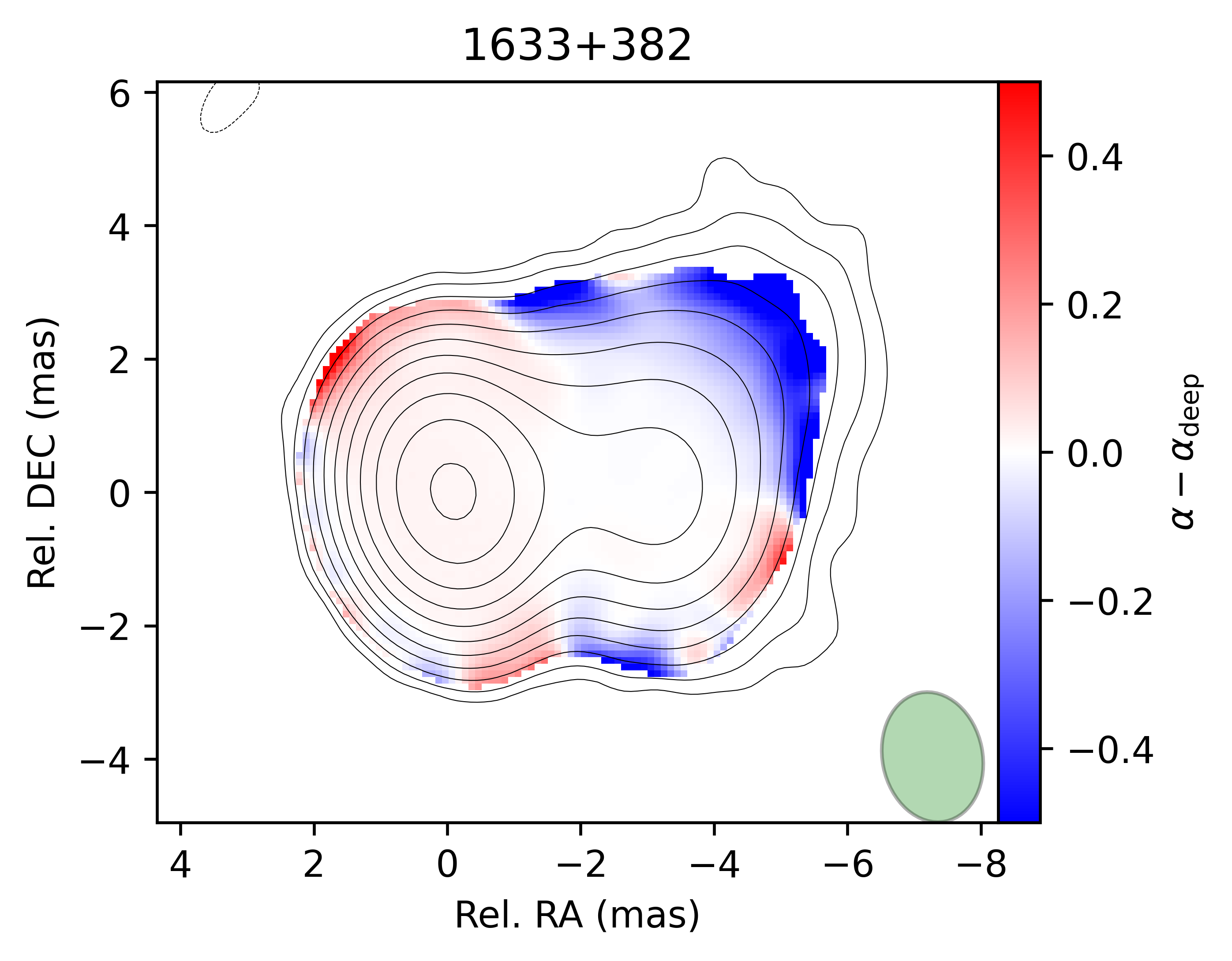}\vspace{0.5cm}
    \includegraphics[width=0.49\linewidth]{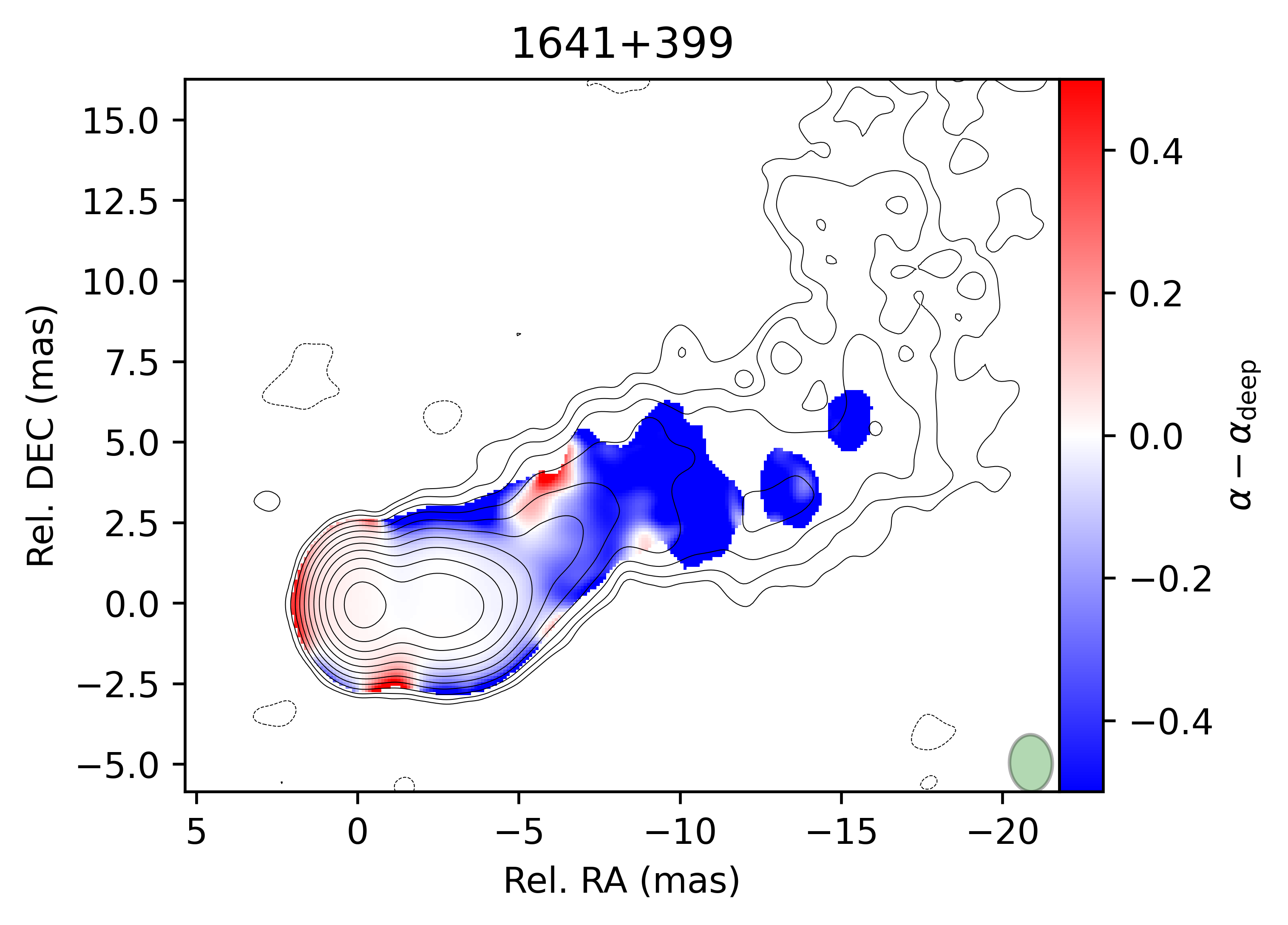}\vspace{0.5cm}
    \caption{Difference between spectral index images made from the conventionally and deep CLEAN Stokes $I$ images (colour). Contours represents 8.1 GHz Stokes $I$ intensity and spectral images are masked according to the conventionally CLEANed Stokes $I$ images. 
    }
    \label{fig:deep_clean_diff}
\end{figure*}

\begin{figure*}
    \centering
    \includegraphics[width=0.49\linewidth]{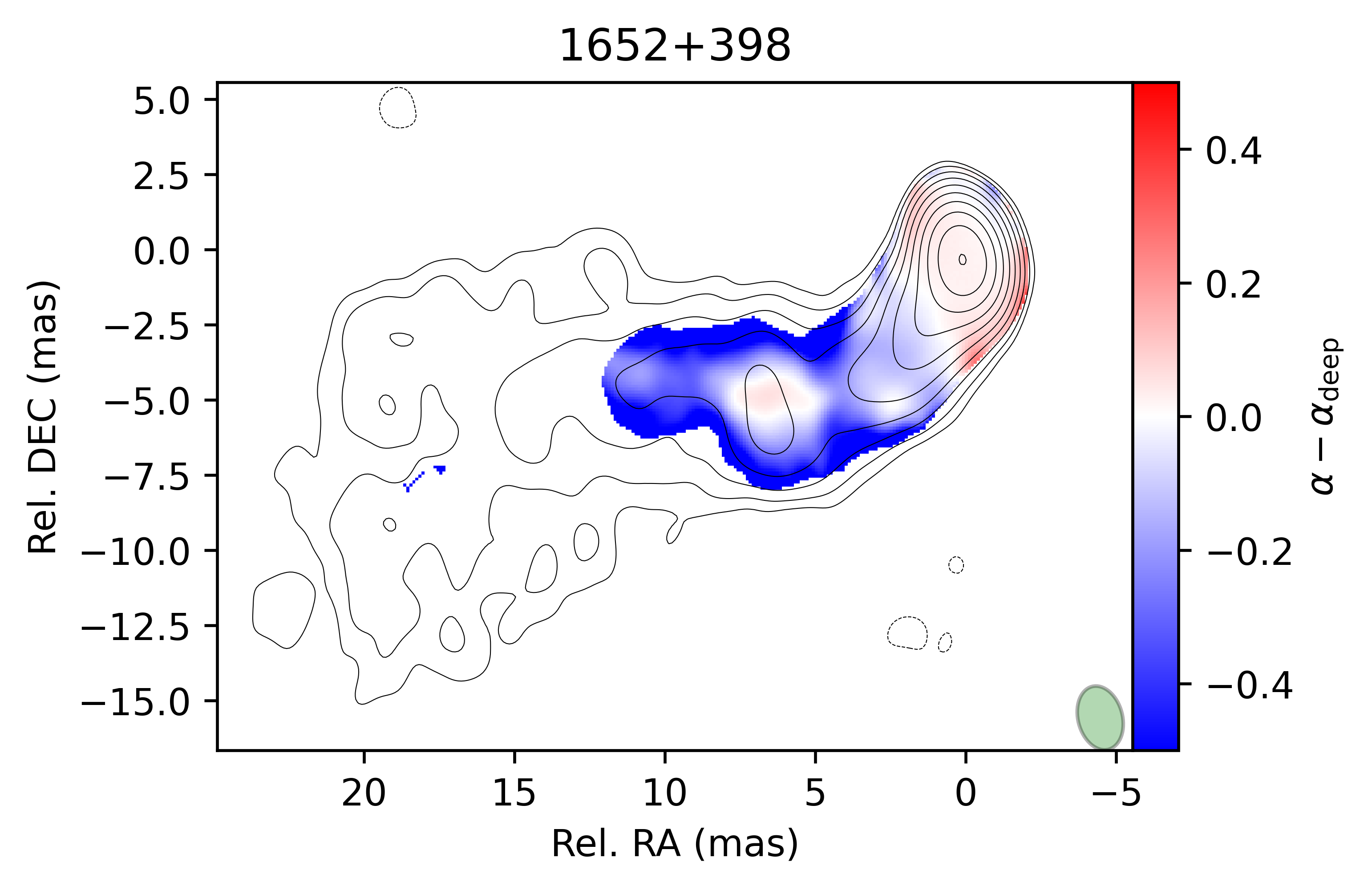}\vspace{0.5cm}
    \includegraphics[width=0.49\linewidth]{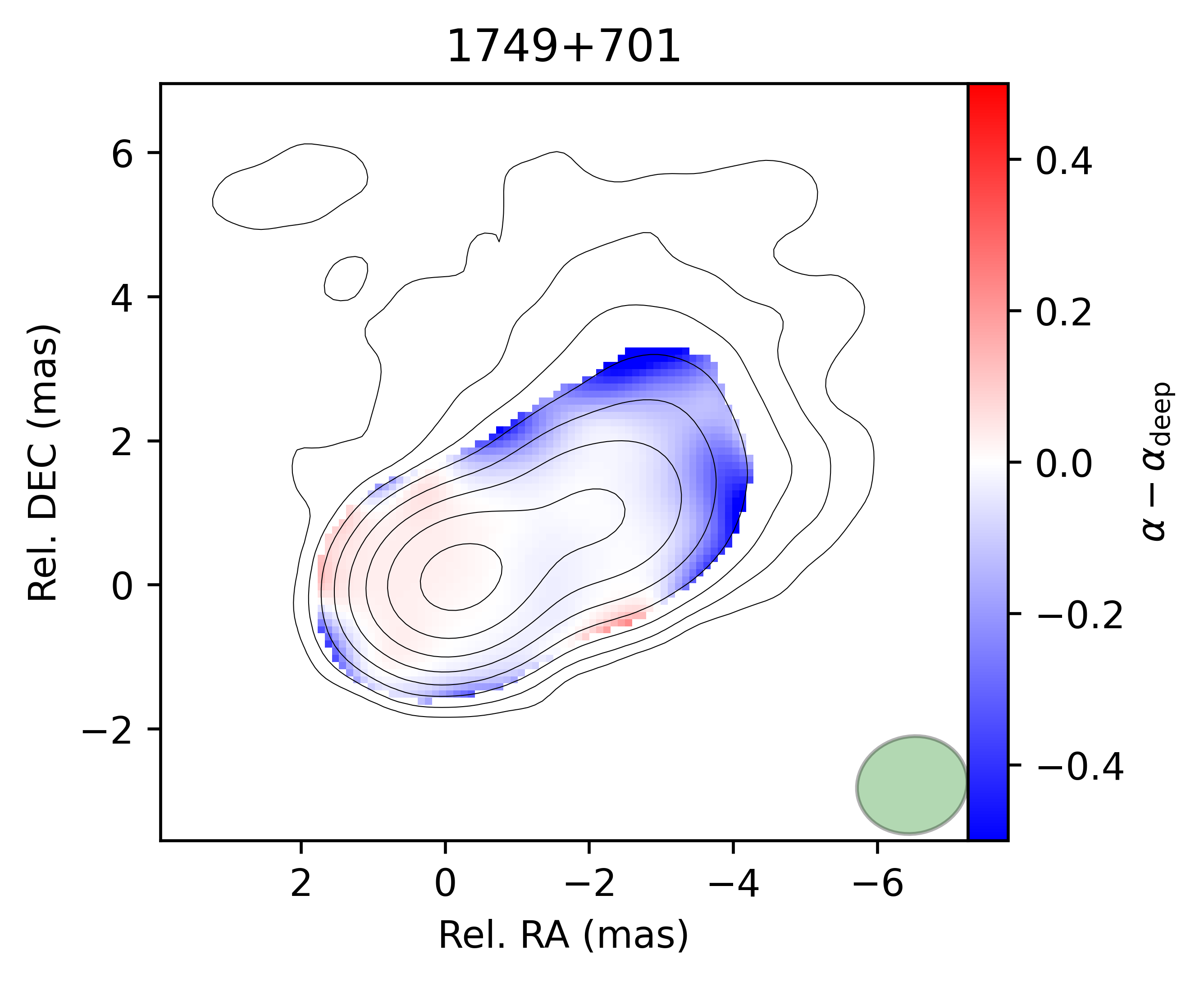}\vspace{0.5cm}
    \includegraphics[width=0.49\linewidth]{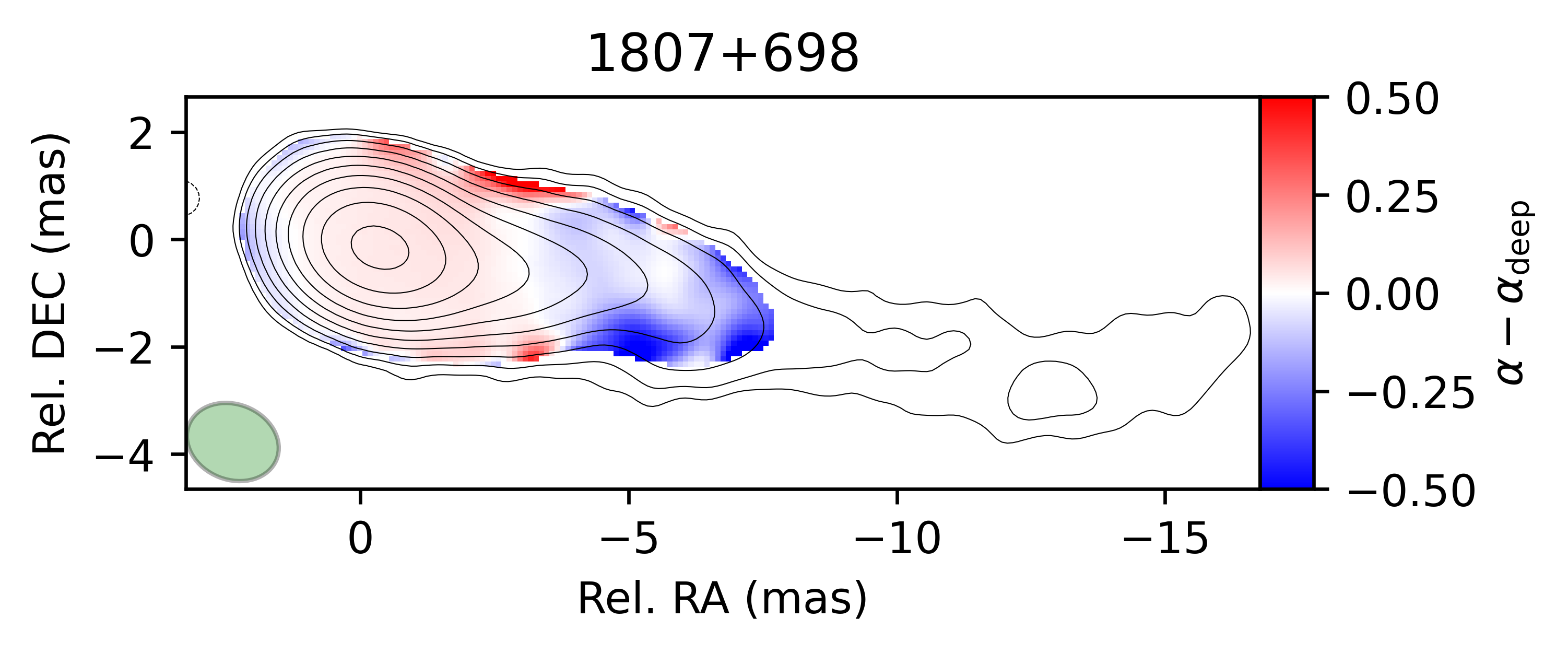}\vspace{0.5cm}
    \includegraphics[width=0.49\linewidth]{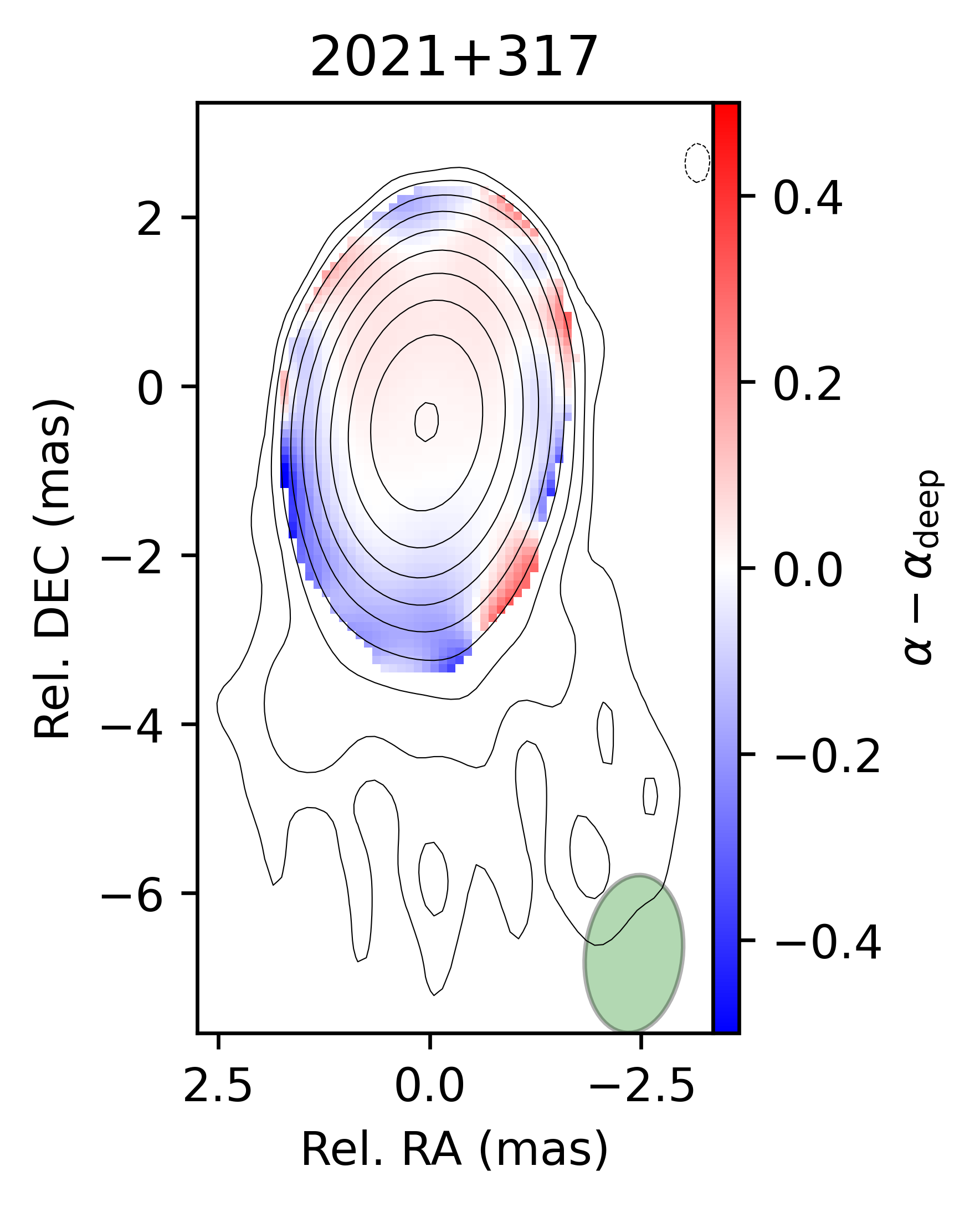}\vspace{0.5cm}
    \contcaption{}
    
    % \label{fig:deep_clean_diff}
\end{figure*}

\end{document}